\documentclass{emulateapj}
\usepackage{natbib}

\slugcomment{To Appear in the AJ}
\shorttitle{SN 2008bk}
\shortauthors{Van Dyk et al.}

\begin{document}

\title{Supernova 2008bk and Its Red Supergiant
  Progenitor}

\author{Schuyler D.~Van Dyk\altaffilmark{1}}
\author{Tim J.~Davidge\altaffilmark{2}}
\author{Nancy Elias-Rosa\altaffilmark{1,3}}
\author{Stefan Taubenberger\altaffilmark{4}}
\author{Weidong Li\altaffilmark{5}}
\author{Emily M.~Levesque\altaffilmark{6,7}}
\author{Stanley Howerton\altaffilmark{8}}
\author{Giuliano Pignata\altaffilmark{9}}
\author{Nidia Morrell\altaffilmark{10}}
\author{Mario Hamuy\altaffilmark{11}}
\author{Alexei V.~Filippenko\altaffilmark{5}}

\altaffiltext{1}{Spitzer Science Center/Caltech, Mailcode 220-6,
  Pasadena CA 91125; email: vandyk@ipac.caltech.edu.}
\altaffiltext{2}{Herzberg Institute of Astrophysics, National Research
  Council of Canada, Victoria, B.C. Canada V9E 2E7; email:
  tim.davidge@nrc.ca.}
\altaffiltext{3}{Institut de Ci\`encies de l'Espai (IEEC-CSIC), Facultat de Ci\`encies, Campus UAB, 08193 Bellaterra, Spain; email: nelias@ieec.cat.}
\altaffiltext{4}{Max-Planck-Institut f\"ur Astrophysik,
  Karl-Schwarzschild-Str.~1, 85741 Garching bei M\"unchen, Germany;
  email: tauben@mpa-garching.mpg.de.}
\altaffiltext{5}{Department of Astronomy, University of California,
  Berkeley, CA 94720-3411; email: weidong@astro.berkeley.edu,
  alex@astro.berkeley.edu.}
\altaffiltext{6}{CASA, Department of Astrophysical and Planetary Sciences, University of Colorado 
389-UCB, Boulder, CO 80309; email: Emily.Levesque@colorado.edu.}
\altaffiltext{7}{Einstein Fellow.}
\altaffiltext{8}{1401 South A, Arkansas City, KS 67005; email: watchingthesky2003@yahoo.com.}
\altaffiltext{9}{Universidad Andres Bello, Departamento de Ciencias
  Fisicas, Avda.~Republica 252, Santiago, Santiago RM, Chile; email:
  gpignata@unab.cl.}
\altaffiltext{10}{Las Campanas Observatory, Carnegie Observatories,
  Casilla 601, La Serena, Chile; email: nmorrell@lco.cl.}
\altaffiltext{11}{Departamento de Astronomia, Universidad de Chile,
  Casilla 36-D, Santiago, Chile; email: mhamuy@das.uchile.cl.}

\begin{abstract}
We have obtained limited photometric and spectroscopic data for
Supernova (SN) 2008bk in NGC 7793, primarily at $\gtrsim 150$ d after explosion. 
We find that it is a Type II-Plateau (II-P) SN which most closely
resembles the low-luminosity SN 1999br in NGC 4900. Given the overall
similarity between the observed light curves and colors of SNe 2008bk
and 1999br, we infer that the total visual extinction to SN 2008bk ($A_V=0.065$ mag)
must be almost entirely due to the Galactic foreground, similar to
what has been assumed for SN 1999br. 
We confirm the
identification of the putative red supergiant (RSG) progenitor star of the
SN in high-quality $g'r'i'$ images we had obtained in 2007 at the
Gemini-South 8 m telescope. Little ambiguity exists in this progenitor
identification, qualifying it as the best example to date,
next to the identification of the star Sk $-$69\arcdeg~202 as the progenitor of SN 1987A.  
From a combination of photometry of the Gemini images with that of archival,
pre-SN, Very Large Telescope $JHK_s$ images, we derive an accurate
observed spectral energy distribution (SED) for the progenitor.  
We find from nebular strong-intensity emission-line indices for several H\,{\sc ii} regions
near the SN that the metallicity in the environment is likely subsolar 
($Z \approx 0.6 {\rm Z}_{\odot}$). 
The observed SED of the star agrees quite well with synthetic SEDs obtained from 
model red supergiant atmospheres with effective temperature $T_{\rm eff}=3600 \pm 50$ K.  
We find, therefore, that the star had a bolometric luminosity with respect to
the Sun of $\log (L_{\rm bol}/{\rm L}_{\odot})=4.57 \pm 0.06$ and
radius $R_{\star} = 496 \pm 34\ {\rm R}_{\odot}$ at $\sim 6$ months prior to explosion. 
Comparing the progenitor's properties with 
theoretical massive-star evolutionary models, we conclude that the
RSG progenitor had an initial mass in the range of 8--8.5 M$_{\sun}$. 
This mass is consistent with, albeit at the low end of, the inferred range of
initial masses for SN II-P progenitors. It is also consistent with the estimated
upper limit on the initial mass of the progenitor of SN 1999br, and it agrees
with the low initial masses found for the RSG progenitors of other low-luminosity SNe~II-P.
\end{abstract}

\keywords{supernovae: general --- supernovae: individual (SN 2008bk)
  --- stars: late-type --- stars: evolution --- stars: fundamental
  parameters: other --- galaxies: individual (NGC 7793)}

\section{Introduction}

The evolution of massive stars, particularly their endpoints, remains
a fascinating, and still unresolved, set of questions and problems
that have ramifications in various aspects of astrophysics, such as
galactic evolution and chemical enrichment.  Theoretically, we expect
stars with initial masses greater than 8--9 M$_{\sun}$ to
terminate central nuclear burning while in the red supergiant (RSG)
phase of their evolution, undergo core collapse, and explode as
supernovae \citep[SNe;][]{woo86,heger03}.  From observations of the progenitor
stars of Type II-Plateau (II-P) SNe, we find that this theoretical lower mass
limit is essentially what is being inferred \citep{sma09}. 
The SNe II-P from detected stars near this limit have tended to be of relatively 
low luminosity, such as SN 2005cs in M51 \citep{maund05,li06} 
and SN 2009md in NGC 3389 \citep{fraser11}. This stands in contrast to 
models which argue that low-luminosity SNe II-P can be explained by
fallback of ejecta on a newly formed black hole, as a result of the core collapse 
of a highly massive ($\gtrsim 20$ M$_{\sun}$) progenitor 
\citep[e.g.,][]{tur98,zam03,nomoto06}.

In recent years, SN progenitors have been 
directly identified in pre-SN archival images obtained by the 
{\sl Hubble Space Telescope} ({\sl HST}).  Of course, direct progenitor
identification has also been made from the ground prior to that, the
most famous example being the discovery of the star Sk $-$69\arcdeg 202
as the progenitor of SN 1987A in the Large Magellanic Cloud
\citep[LMC; e.g.,][]{gil87,son87}.  Other cases include the
progenitors of SN 1961V in NGC 1058 \citep{ber64, zwi64}, SN 1978K in
NGC 1313 \citep{ryd93}, and SN 1993J in M81
\citep{ald94,coh95,van02}.  However, all of these examples are
SNe which in some way have been peculiar or otherwise unusual.  The
nature of SN 1961V as a true SN continues to be debated (see \citealt{van05}, and
references therein; also, \citealt{smith11}; \citealt{kochanek11}; \citealt{van11}).  

\begin{figure}
\includegraphics[angle=0,scale=0.70]{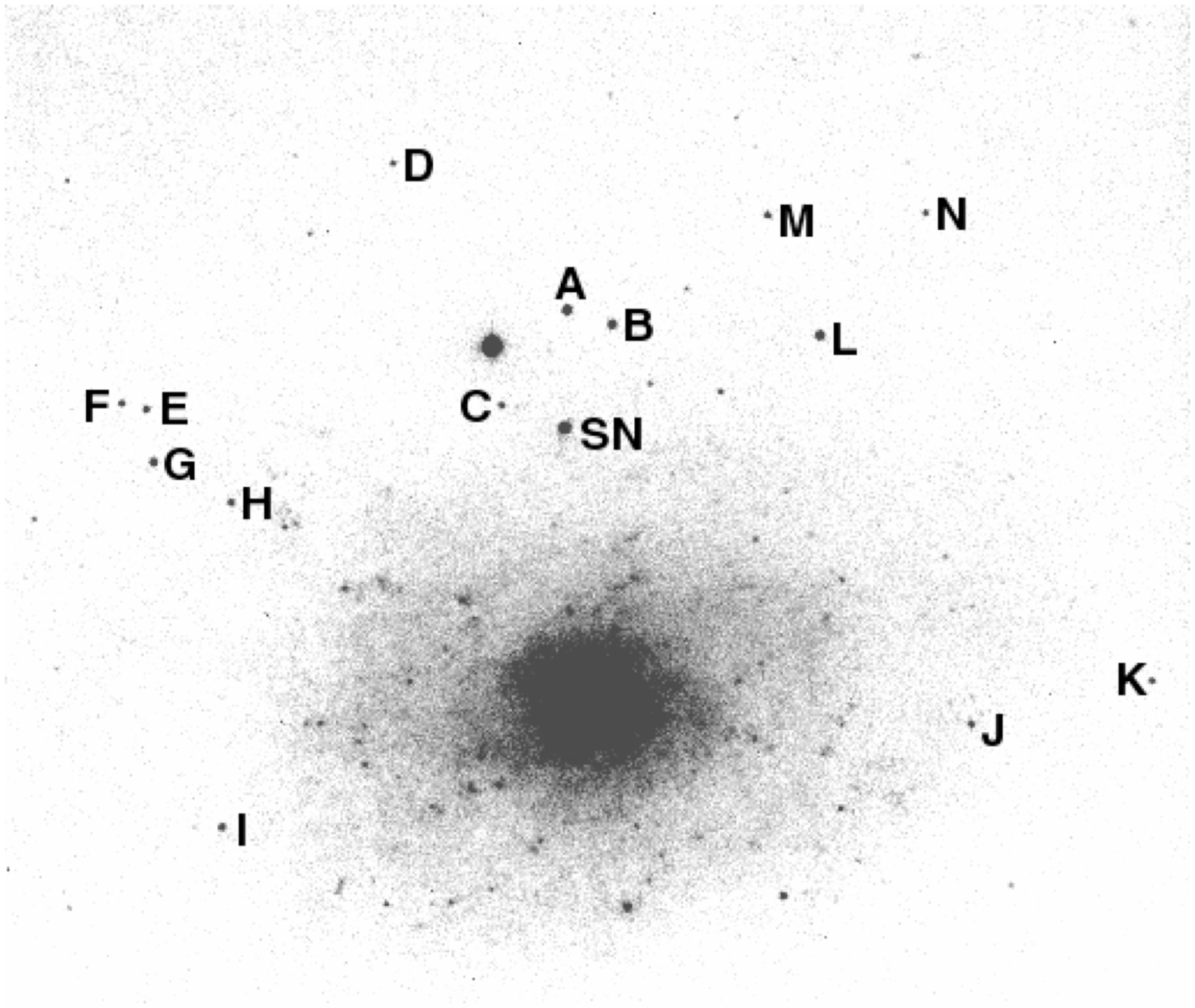}
\caption{Chart of SN 2008bk in NGC 7793 in an $I$-band image obtained
  with the Calar Alto 2.2 m telescope on 2008 September 13.  The
  comparison stars from Table \ref{tabfids}, as well as the SN, are
  labeled.  North is up and east is to the left.\label{figchart}}
\end{figure}

SN 2008bk in the nearby, nearly face-on spiral galaxy NGC 7793 (see
Figure~\ref{figchart}) has also had its progenitor star identified in 
ground-based data.  The SN was discovered by \cite{mon08} on March 25.14 (UT dates
are used throughout this paper).  \cite{li08} were the first to
attempt to identify the progenitor star as a likely RSG in deep
archival ground-based $BVI$ images obtained in 2001 with one of the
8.2 m Unit Telescopes (UT) of the European Southern Observatory (ESO) Very
Large Telescope (VLT). They used the Monard discovery images and also
images taken on March 28.41 with the 0.41 m PROMPT 2 telescope at
the Cerro Tololo Inter-American Observatory (CTIO) as references.
This search for the progenitor was attempted even before the SN had
been classified. However, Li et al.~concluded that SN 2008bk was likely a
core-collapse SN, since a star was detected at the location; it is far
less likely that we would be able to detect the progenitor, or the
companion to the progenitor, of a thermonuclear Type Ia SN outside of
the Local Group \citep[see, e.g.,][]{li11}. The SN was subsequently classified by \citet{mor08}
as Type II with an age of 36~d after explosion on April 12.4, based on
a comparison with the well-studied SN II-P 1999em in NGC 1637.

\citet{maoz08} detected and measured magnitudes for the progenitor
star in archival near-infrared (IR) $J$- and $K_s$-band images, also obtained
with the VLT in 2005.  \citet{mat08} further analyzed the VLT optical
images, as well as VLT $H$-band data obtained in 2007, and combined
these pre-SN results with precise astrometry provided by post-SN $K_s$
VLT adaptive optics imaging. From their analysis they concluded that
the total extinction toward the progenitor is $A_V=1.0 \pm 0.5$ mag,
and that the star therefore was likely of type M4~I, with an initial
mass of $8.5 \pm 1.0\ {\rm M}_{\sun}$.

In this paper, with limited photometry and spectroscopy of the SN
itself, we show that SN 2008bk is a SN~II-P very similar to the
low-luminosity SN 1999br in NGC 4900 \citep{pas04}. 
Furthermore, we
measure accurate photometry for the RSG progenitor from
superior-quality $g'r'i'$ images obtained in 2007 with
the Gemini-South 8 m telescope, as well as from
the archival, pre-SN, near-IR VLT images.
We compare this photometry to spectral energy distributions (SEDs) obtained
from stellar atmosphere models for RSGs, to infer the star's effective
temperature. Together with the inferred luminosity for the star, 
we compare these
properties to recent theoretical models for massive-star evolution,
to estimate the initial mass for the progenitor.
We also compare our results to those obtained by \citet{mat08}.

\section{Supernova Observations}

Although highly incomplete in terms of photometric and spectroscopic
coverage, our observations of SN 2008bk are still sufficient to
determine its general properties and to allow for comparison with
other SNe~II.  A more complete analysis of the properties of SN 2008bk
is forthcoming (Pignata et al., in prep.).

\subsection{Photometry}\label{photom}

We obtained $BVRI$ images of SN 2008bk, at eight epochs late
in its evolution: four at the Calar Alto 2.2 m telescope + the CAFOS
SiTe (scale $0{\farcs}53$ pix$^{-1}$, field of view $9\arcmin \times
9\arcmin$), two at the 2.5 m Nordic Optical Telescope (NOT; Roque de
los Muchachos Observatory, La Palma, Spain) + ALFOSC ($0{\farcs}19$
pix$^{-1}$, field of view $6{\farcm}5 \times 6{\farcm}5$), and two
using the SMARTS 1.3 m telescope + ANDICAM at CTIO. These images were
processed using standard IRAF\footnote{IRAF (Image Reduction and 
Analysis Facility) is distributed by the National Optical Astronomy 
Observatories, which are operated by the Association of Universities 
for Research in Astronomy, Inc., under cooperative agreement with 
the National Science Foundation (NSF).} procedures for image trimming, bias
subtraction, and flat-field corrections.  Additionally, we add several
relatively early-time epochs at $V$, with an image obtained at the Las
Campanas Observatory (LCO) 2.5 m duPont telescope on 2008 April 13
and six images obtained with the Global Rent-A-Scope (GRAS) 0.32 m and 
0.41 m telescopes at epochs spanning the plateau to early on the
exponential tail.  We have also included $BV$ data, obtained by the
CHilean Automatic Supernova sEarch \citep[CHASE;][]{pig08} with PROMPT 2.

We extracted instrumental photometry for the SN and
reference stars from these images, generally, via point-spread function (PSF)
fitting using DAOPHOT \citep{ste87} within IRAF. When the shapes of
the stars in images deviated substantially from a ``well-behaved''
PSF, we relied only on aperture photometry.  For this reason, we
calibrated the SN and reference star instrumental photometry from all
of the images to existing photometry for the star field around the
host galaxy.  NGC 7793 was observed by \citet{lar99b} in $UBVRI$ with
the Danish 1.54 m telescope + the Danish Faint Object Spectrograph and
Camera on 1997 September 6.  We obtained these images, posted for
public distribution in FITS format, from the NASA/IPAC Extragalactic
Database (NED\footnote{http://nedwww.ipac.caltech.edu.}).  We
extracted instrumental magnitudes from the images using IRAF, first
with a 4\arcsec\ aperture and then with a PSF for each band.
Calibration was established using photometry of a number of isolated
reference stars through a 20\arcsec\ aperture, which matches the aperture 
used to establish the calibration through standard-star observations by
\citet[his Table 1]{lar99a}; see Figure~\ref{figchart} and 
Table~\ref{tabfids} for the locations and magnitudes of these stars. 
We also applied the appropriate extinction corrections (S.~S.~Larsen
2010, private communication).

We can compare our $BVR$ photometry with those magnitudes provided in
NOMAD\footnote{http://www.nofs.navy.mil/nomad.html.}; however, the
uncertainties in these latter magnitudes are substantial, resulting in
large scatter in the comparison: for $B$, NOMAD$-$Ours $= -0.32 \pm
0.20$; for $V$, $-0.51 \pm 0.15$; and for $R$, $-0.39 \pm 0.19$. A
far better comparison can be made at $I$ with the more reliable
photometry from the DENIS database \citep{denis}: DENIS$-$Ours $=
0.02 \pm 0.05$ mag.  The good agreement in the $I$ magnitudes
provides us with some confidence, overall, in the calibration of our
photometry.

As many reference stars as possible were
generally adopted for the calibration; however, for the shallowest
images, we had to use a combination of the magnitudes of star K (the
brightest star in Table~\ref{tabfids}) and the less precise $BV$
photometry from the bright NOMAD stars seen in these images.  The
resulting photometry of SN 2008bk is given in Table~\ref{tabphot}.
The uncertainties in the SN photometry in all four bands arise from
the measurement uncertainties from DAOPHOT, the uncertainties in the
absolute calibration given by \citet{lar99a}, and the uncertainties in
transforming the instrumental magnitudes to standard magnitudes, based
on the reference stars, all added in quadrature.

The $BVRI$ light curves for SN 2008bk are shown in Figure~\ref{figlc}.
As we have noted, many of the multi-band points are beyond the plateau
phase and on the exponential tail. However, the $V$ photometry at
earlier times provides us with adequate leverage on the overall
photometric evolution of SN 2008bk, from soon after explosion through
the onset of this tail. The lone early-time $B-V$ datapoint provides us
with some leverage on the initial color of the SN.

For comparison, we also show in Figure~\ref{figlc} the $V$ light
curves for both the ``normal'' SN II-P 1999em \citep{hamuy01,leo02,elm03}
and the subluminous SN II-P 2005cs in M51 \citep{pas09}, and,
for all bands, the subluminous SN II-P 1999br \citep{pas04}.
We have attempted to match, by eye, the brightnesses of these
comparison SNe with SN 2008bk toward the end of the plateau phase at $V$.  
Although we have very limited coverage at the end of the plateau, it
appears that this phase ended very near JD 2454651 (i.e., 2008 July 3), implying
that the plateau lasted $\sim 100$ d after discovery.
(Additionally, the end of the plateau phase for SN 1999br is not well defined;
\citeauthor{pas04} \citeyear{pas04}.)
The SN 2008bk light curves are shown in
Figure~\ref{figlc} relative to that epoch. 

\begin{figure}
\includegraphics[angle=0,scale=0.70]{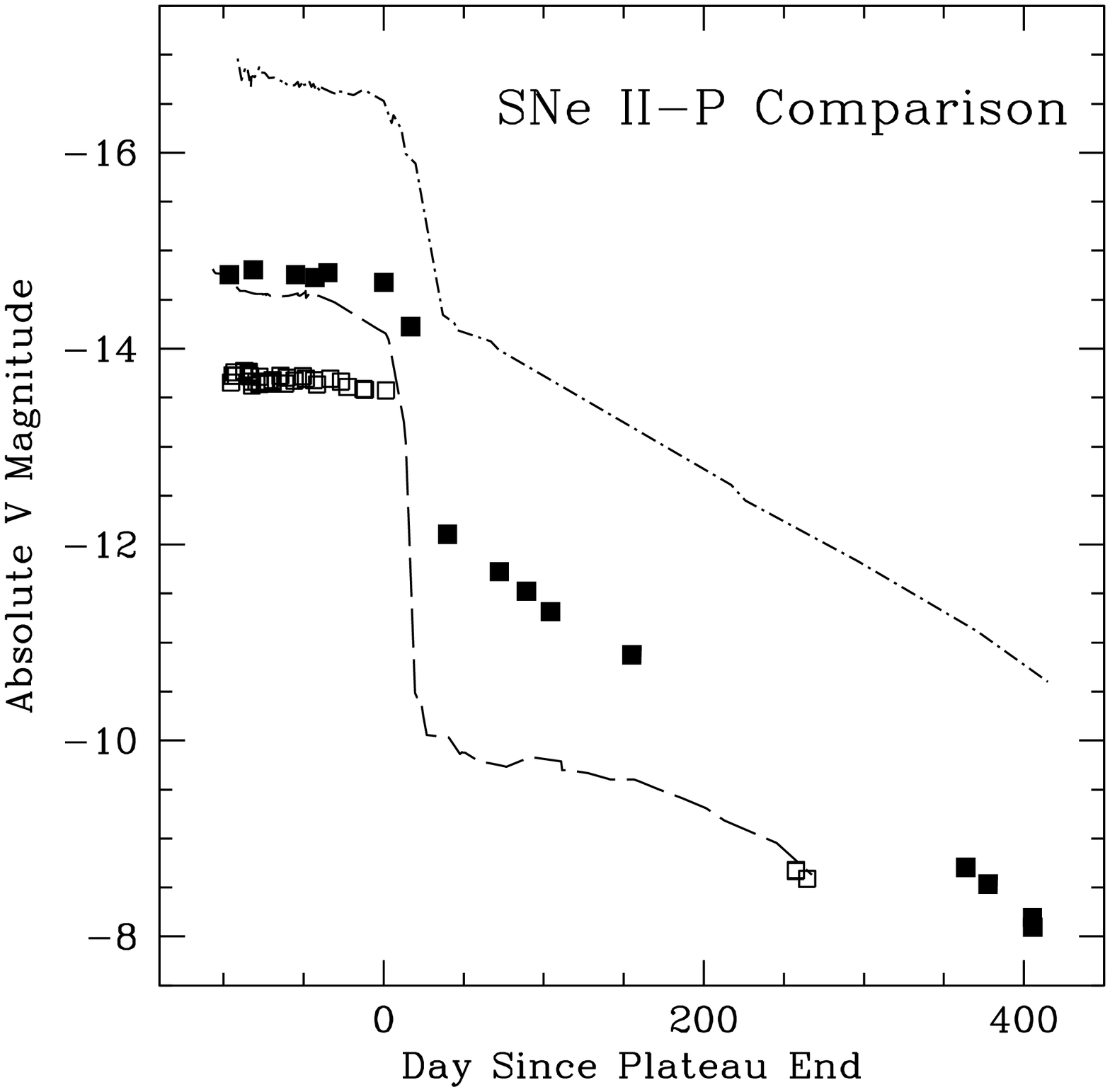}
\caption{The evolution of the absolute $V$ magnitude of SN 2008bk ({\it solid points}).  
  Shown for comparison are the absolute $V$ light curves for the subluminous SN II-P 1999br
  in NGC 4900 \citep[{\it open points};][]{pas04}, as well as for the SNe II-P 1999em in 
  NGC 1637 ({\it dot-dashed line}; \citealt{hamuy01,leo02,elm03}) and 2005cs in M51 
  ({\it dashed line}; \citealt{pas09}). The observed $V$ light curves for the SNe have been
  corrected for their respective extinction and distance; see text.
  All light curves are shown
  relative to the epoch corresponding to the end of the plateau phase.\label{figlcabs}}
\end{figure}

In Figure~\ref{figlcabs} we show the absolute $V$ light curve for SN 2008bk,
which results from correcting for the assumed extinction (see \S~\ref{extinction})
and distance modulus (see \S~\ref{distance}) to the SN.
We also show for comparison the absolute $V$ light curves for
SN 1999em, corrected for the total extinction, $A_V=0.31$ mag \citep[e.g., ][]{hamuy01},
and the Cepheid distance modulus to its host galaxy \citep{leo03}; 
SN 2005cs, corrected for the extinction from \citet{pas09} and 
the distance modulus from \citet{tonry01}; and
SN 1999br (extinction and distance modulus from \citeauthor{pas04}~\citeyear{pas04}).
We estimate that SN 2008bk had an absolute
magnitude near maximum brightness during the plateau phase of
$M^0_V=-14.8$ mag, which is $\sim 2.1$ mag less luminous than SN 1999em.
Although the plateau brightness for SN 2008bk is very similar to the $M^0_V=-14.9$ mag for the 
low-luminosity SN~II-P 2005cs \citep{pas09}, SN 2008bk
did not decline nearly as rapidly, post-maximum, nor is it nearly as subluminous 
on the tail as SN 2005cs. 
One can see from Figure~\ref{figlc} that SN 2008bk exhibits a
striking photometric resemblance to SN 1999br at all bands. 
However, a noticeable offset exists
between the absolute brightnesses of SNe 2008bk and 1999br; we comment on this more
in \S~\ref{conclusions}.
We 
will demonstrate in \S \ref{spec} that SN 2008bk also spectroscopically most
closely resembles SN 1999br.  We therefore consider SN 2008bk to be a 
low-luminosity SN~II-P, quite similar to SN 1999br.

The photometric comparison with SN 1999br \citep[assuming the 
explosion date for SN 1999br is JD 2451278;][]{pas04} indicates that
the explosion epoch for SN 2008bk was $\sim$ JD 2454550, or 
$\sim$ 2008 March 24.  Monard, therefore, likely discovered the
SN $\lesssim 1$~d after explosion.

\subsection{Spectroscopy}\label{spec}

We obtained several spectra of SN 2008bk at late times, using the
Blanco 4 m telescope at CTIO with the RC-Spectrograph in 2008 July and
using the Calar Alto 2.2 m telescope with the CAFOS SiTe imaging
spectrograph in 2008 September.  The details of these observations are
given in Table~\ref{tabspec}.  The spectra were reduced using standard
IRAF routines. The preprocessing of the
spectroscopic data (trimming, bias, overscan, and flat-field
correction) were the same as for the images (see \S \ref{photom}). The
one-dimensional spectra were wavelength calibrated by comparison with
arc-lamp spectra obtained during the same night and with the same
instrumental configuration, and flux calibrated using
spectrophotometric standard stars. The zeropoint of the wavelength
calibration was verified against the bright night-sky emission
lines. The standard-star spectra were also used to model and remove
the telluric absorption \citep[e.g.,][]{math00}.

\begin{figure}
\includegraphics[angle=0,scale=0.70]{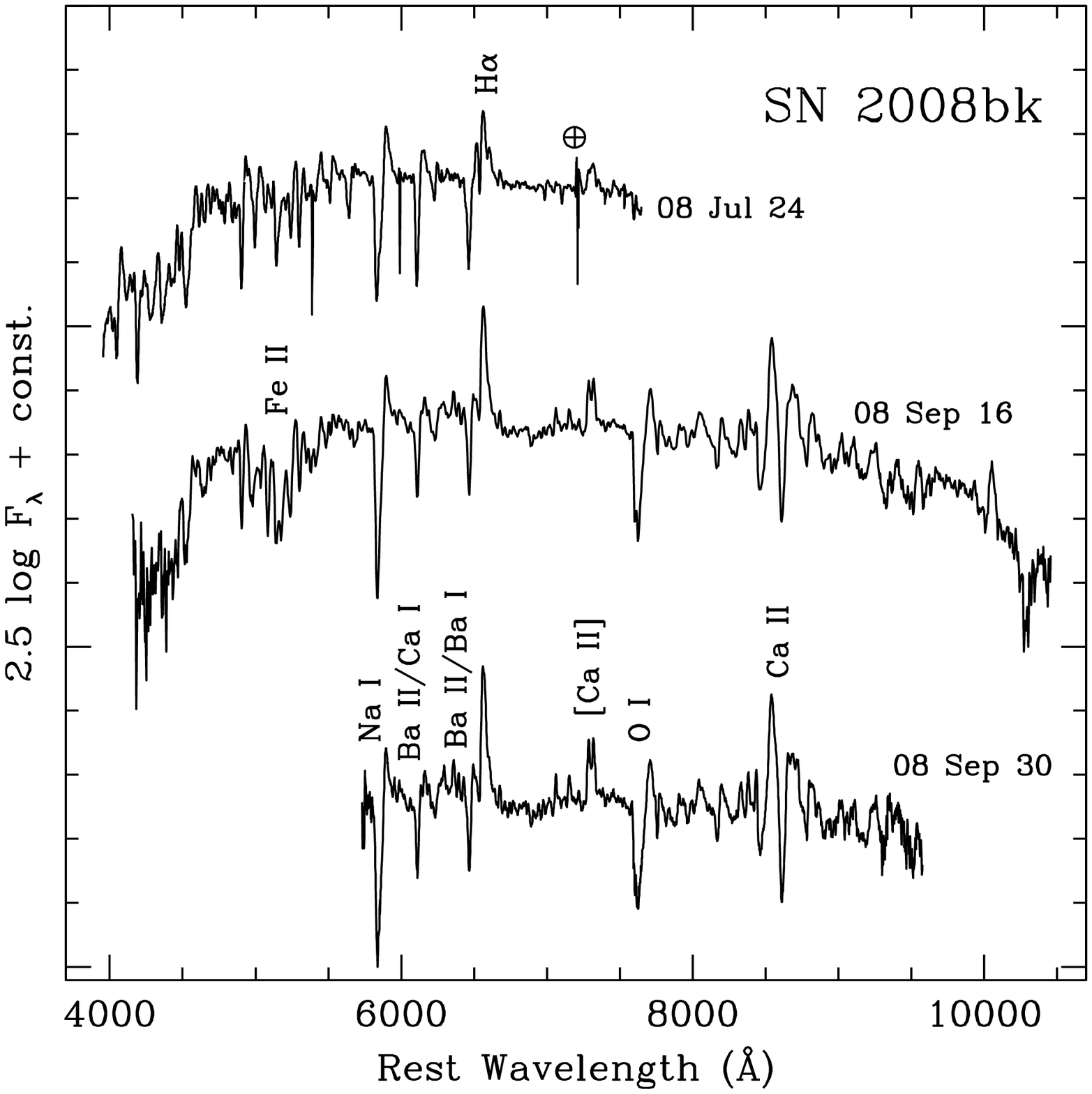}
\caption{Optical spectra of SN 2008bk obtained with the 4 m CTIO 
  Blanco telescope on 2008 July 24, 
  and with the 2.2 m Calar Alto telescope on 2008 September 16 and 30. 
  See Table~\ref{tabspec} and the text. 
  Several notable features in the spectra are
  indicated.\label{figspec1}}
\end{figure}

\begin{figure}
\includegraphics[angle=0,scale=0.70]{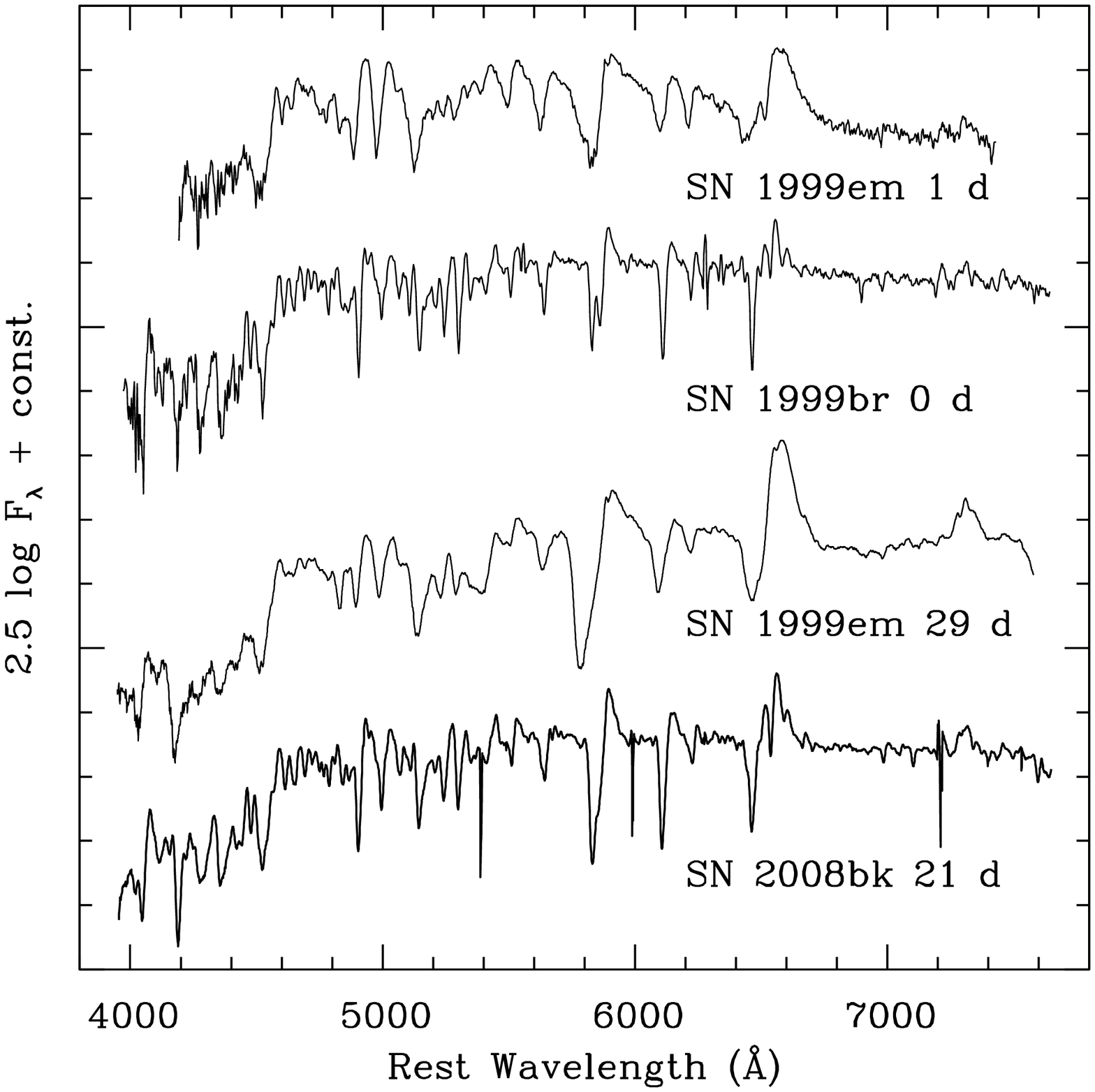}
\caption{A comparison of the SN 2008bk optical spectrum from 2008 July
  24 (see Table~\ref{tabspec}) with the spectra of SN 1999em 
  \citep{leo02} and a spectrum of SN 1999br 
  \citep{pas04}. The ages, in days, for the SNe in the figure are relative to the
  estimated epoch for the end of the photometric plateau phase in each case.\label{figspec2}}
\end{figure}

The resulting spectra are shown in Figure~\ref{figspec1}.  Since the
spectra obtained at CTIO are only spaced in time each by a day, the
features in the spectra had varied little in that time interval.
Thus, we show only a representative spectrum from July 24. The same
is true for the two spectra obtained at Calar Alto in mid-September;
we display only the spectrum from September 16. 
The July 24 spectrum is $\sim 21$ d past the plateau phase, while the
September 16 and 30 spectra are during the nebular phase, 
at $\sim 75$ and $\sim 89$ d past the plateau.
In Figure~\ref{figspec2} we show the SN 2008bk spectrum from July
24, and also spectra of the normal SN II-P 1999em \citep{leo02} 
at 1~d 
and 29~d 
past the plateau phase, 
as well as a spectrum of the subluminous SN II-P 1999br in
NGC 4900 \citep{pas04} at essentially the end of the plateau. 
The spectra for SN 1999em were obtained from the SUSPECT
SN database\footnote{http://suspect.nhn.ou.edu/$\sim$suspect.}.  
A.~Pastorello kindly provided the spectrum from the Padova SN data archive
of SN 1999br from 1999 July 20. 

Both the strength and width of
the various spectral lines, and the overall continuum shape, for SN 2008bk
compare rather poorly with SN 1999em.
A far better comparison, overall, can be made with SN 1999br.
The relative strength of, for
example, the [Ca\,{\sc ii}] $\lambda\lambda$7291, 7324 feature in the
SN 2008bk spectrum, compared with that in the SN 1999br spectrum, again 
indicate that SN 2008bk is also a
low-luminosity SN II-P. 

\section{Detection of the Progenitor Star}\label{detection}

\citet{li08} measured a precise position for the SN of $\alpha$(J2000)
= $23^{\rm h}57^{\rm m}50{\fs}42$, $\delta$(J2000) = $-32\arcdeg
33\arcmin 21{\farcs}5$.  Within the uncertainty of this position ($\pm
0{\farcs}2$) they isolated a star in the VLT $I$-band image at this
location.  The star is not detected in the $B$ and $V$ images from the
same epoch.  This candidate progenitor lies within a prominent cluster
of bright blue and red supergiant stars.  From its environment and its
lack of detection in the bluer bands, Li et al.~conjectured that the
progenitor was a RSG.

We have since improved the astrometric agreement, with an uncertainty
of $0{\farcs}13$, between the SN and the star with a $V$ image of the
SN from the LCO 2.5 m duPont telescope, obtained on 2008 April 13.  We
have refined the absolute position for the SN to end figures
$50{\fs}50$ and $20{\farcs}7$ (with uncertainty $\pm 0{\farcs}2$).

We obtained deep $g'r'i'$ images in 2007 September and October with
GMOS (Crampton et al. 2000) on Gemini-South, to study the outer disk
of NGC 7793. The images were obtained in queue observing mode as part
of program GS2007B-Q-57 (PI: Davidge). The detector in GMOS is a mosaic of three
CCDs with a raw image scale of $0{\farcs}0727$ pix$^{-1}$. The data
were binned $2 \times 2$ pixels$^2$ to better match the delivered
image quality. Eight 200-s exposures were obtained in $g'$, and eight
300-s exposures were obtained in both $r'$ and $i'$. The raw images in
each band were bias subtracted, divided by a twilight flat, and then
stacked.  The progenitor star and its environment are shown in
Figure~\ref{figimage}. From these images we measure an absolute
position for the star to end figures $50{\fs}48$ and $20{\farcs}9$
($\pm 0{\farcs}2$).  To within the absolute astrometric uncertainties,
the position of the identified progenitor star agrees very well with
the SN position (see above).

We extracted instrumental photometry from the $g'r'i'$ images via PSF
fitting within IRAF/DAOPHOT.  Since the Gemini program was intended to
be carried out during nonphotometric conditions as part of the
so-called ``Poor Weather Queue,'' this photometry must be calibrated
using other observations.  Following Welch et al.~(2007), we then
transformed the $g'r'i'$ instrumental magnitudes from the Gemini
images to the calibrated $VRI$ magnitudes from the Danish telescope
images (see \S \ref{photom}), using a number of bright ($V=16.1$--20.4
mag), isolated stars common to both sets of images.  The resulting
transformations are

\begin{displaymath}
V = 0.7803 g' - 0.0181 r' + 0.2768 i' + {\rm zeropoint},
\end{displaymath}
\begin{displaymath}
R = 0.0000 g' + 0.7762 r' + 0.2419 i' + {\rm zeropoint},
\end{displaymath}
\begin{displaymath}
I = 0.0000 g' - 0.0722 r' + 1.1055 i' + {\rm zeropoint}.
\end{displaymath}

\noindent We note that the coefficients in the transformation above
are similar to the ones derived by Welch et al.~as applied to a
different scientific context. The uncertainties in these
transformations are 0.08 mag for $V$, 0.05 mag for $R$, and 0.07 mag
for $I$.

The resulting $VRI$ magnitudes for the SN 2008bk progenitor from the
Gemini images are given in Table~\ref{tabprog}. The uncertainties in
the magnitudes include the measurement uncertainties from DAOPHOT, the
uncertainties in the calibration of the Danish telescope images of the
galaxy \citep[][his Table 1]{lar99a}, and the uncertainties in the
transformations above, all added in quadrature.

\bibpunct[; ]{(}{)}{;}{a}{}{;}
We also reanalyzed and remeasured the photometry from the near-IR VLT
ISAAC and HAWK-I $JHK_s$ images, initially analyzed by \citet{maoz08}
and \citet{mat08}. Photometry was performed, again using PSF fitting
with IRAF/DAOPHOT, and the instrumental magnitudes were calibrated to
2MASS stars in the fields. The resulting photometry is given in
Table~\ref{tabprog}.  The uncertainties in the magnitudes include the
measurement uncertainties from DAOPHOT, the uncertainties in the 2MASS
star magnitudes, and the uncertainties in the transformations using
2MASS. Later (see \S~\ref{mass}), we will need these near-IR magnitudes in the 
\citet{bessell88} system, particularly for $K$. For this reason, we use the
transformations from the 2MASS system in \citet[][his Appendix A]{carp01}
to obtain $J=19.34$, $H=18.56$, and $K=18.18$ mag (formal uncertainties in
these transformations are all $< 0.01$ mag).

\section{Properties of the Progenitor Star and Its Environment}\label{properties}

We will now place the SN 2008bk progenitor in a Hertzsprung-Russell
diagram (HRD), to attempt to estimate its initial mass.  
Whereas the photometry of the progenitor star has relatively small uncertainties, 
the next steps, discussed below, involved in analyzing this photometry possess significantly 
larger uncertainties.

\subsection{Extinction to the Supernova}\label{extinction}

Here we estimate the extinction to the SN.
The startlingly good agreement, seen in Figure~\ref{figlc}, between
the light curves of both SN 2008bk and SN 1999br in all observed bands
(the SN 1999br curves have all been uniformly adjusted by $-4.6$ mag
to match, by eye, those of SN 2008bk) implies that the extinction toward
SN 2008bk is quite similar to that of SN 1999br.  \citet{sma09}
assume only the Galactic foreground contribution to the extinction toward
SN 1999br: $A_V=0.065$ mag. We can then logically infer that the
extinction toward SN 2008bk is comparably quite low; we assume
$A_V=0.065$ mag.  Interestingly, the estimate
for the Galactic line-of-sight extinction toward the SN is, in fact,
$A_V=0.065$ mag \citep{sch98}, which would imply that the extinction
in the immediate galactic environment of SN 2008bk is negligible. From
a preliminary analysis of the photospheric spectra for SN 2008bk,
no sign exists of Na\,{\sc i}~D absorption due to the host galaxy.
Additionally, the average internal extinction in NGC 7793 is generally
quite low \citep[$A_B=0.12$ mag;][]{pie92}. The low extinction value
is consistent with the lack of emission at 8.0 $\mu$m (normally
attributed to emission from polycyclic aromatic hydrocarbons in
interstellar dust clouds) detected at the position of SN 2008bk in
archival images of NGC 7793 obtained using the InfraRed Array Camera
(IRAC) on the {\sl Spitzer Space Telescope}.  It is also consistent
with the fact that the SN is within what appears to be a nebular,
possibly evacuated, bubble (see Figure~\ref{fighiiregion}).

\subsection{Distance Modulus to the Host Galaxy}\label{distance}

We also require an estimate of the distance modulus to the SN. \citet{puc88}
calculated a mean distance modulus, $27.64 \pm 0.19$ mag, to NGC 7793 
from several distance indicators, including the Tully-Fisher relation.
\citet{kar03} measured a distance modulus of $27.96 \pm 0.24$ Mpc
(somewhat more distant, but in good agreement with the Puche \&
Carignan value), using {\sl HST\/} images of the host galaxy to estimate 
the ``tip of the red giant branch'' (TRGB) brightness to be $I=23.95 \pm 0.22$ mag. 
(These {\sl HST\/} images, unfortunately, did not cover the SN site; furthermore, we
note that \citealt{mou08}, using the TRGB calibration of $M_I^0=-4.05$
mag by \citealt{riz07}, found $\mu^0=27.78$ mag; however, their
assumed $I$-band extinction, $A_I=0.22$ mag, is almost certainly too
large.) 

The distance modulus has been recently determined using Cepheids in the
host galaxy by \citet{piet10} to be $\mu^0=27.68 \pm 0.05$ mag (random)
$\pm 0.08$ mag (systematic). Those authors note that their distance
modulus also agrees very well with a recent TRGB measurement,
$\mu^0=27.72 \pm 0.08$ mag, by \citet{jacobs09}. Hereafter, we adopt
the Cepheid distance modulus and, conservatively, assume the (larger) 
systematic uncertainty in that estimate.

\subsection{Metallicity of the Supernova Environment}\label{metals}

Furthermore, we need to know the metallicity in the SN environment, 
since we are comparing the progenitor's observed characteristics 
with estimates from models which are generated assuming a particular
metallicity.

\begin{figure}
\includegraphics[angle=0,scale=0.70]{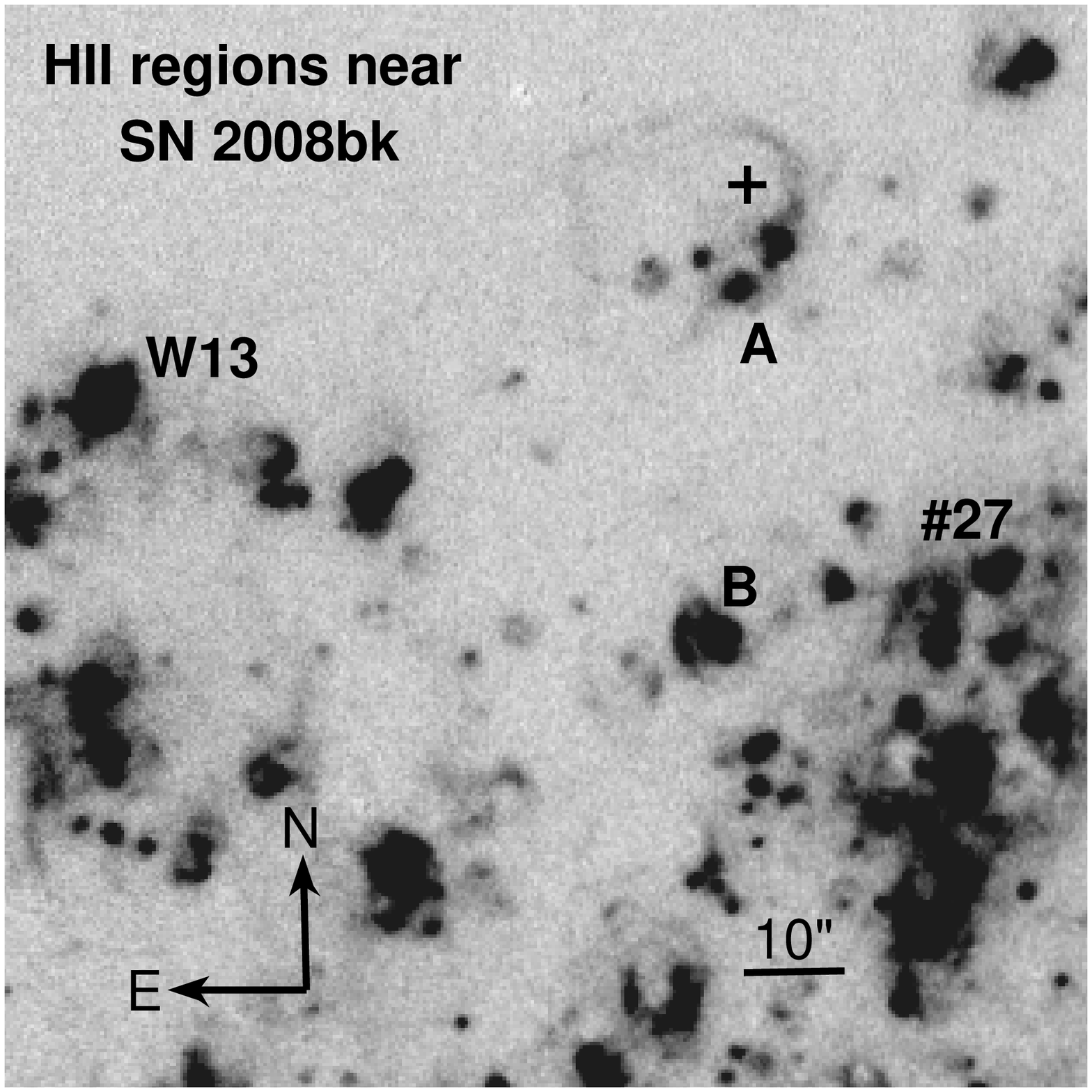}
\caption{The continuum-subtracted H$\alpha$ (+[N\,{\sc ii}]) image of
  NGC 7793, an archival ancillary data product released by the {\sl Spitzer\/}
  SINGS project \citep{kenn03}, with the position of SN 2008bk
  indicated by a {\it cross}. The locations of H\,{\sc ii} regions near the
  SN are labelled. Regions ``A'' and ``B'' were also detected in a spectrum of SN 2008bk
  we obtained with the 2.2 m Calar Alto telescope on 2008 September 30. 
  Region ``\#27'' is from \citet{bibby10}.
  Region W13 is from \citet{mccall85}. See Table~\ref{hiiregions}.
  Note that the SN site appears to be within a faint,
  but extended, bubble of ionized gas, of diameter $\sim 530$ pc. 
  North is up and east is to the left.
\label{fighiiregion}}
\end{figure}

We consider the metallicity to be equivalent to the local oxygen abundance. 
The most accurate estimate of the
metallicity ideally would be from the immediate SN environment
itself, or failing that, from regions of the host galaxy in the general vicinity of the
SN. 
We show several such H\,{\sc ii} regions in Figure~\ref{fighiiregion}, which is 
a continuum-subtracted H$\alpha$ (+ [N\,{\sc ii}]) image
of NGC 7793, released as an ancillary data product by the 
{\sl Spitzer\/} SINGS Legacy project\footnote{Available at
  http://irsa.ipac.caltech.edu.} \citep{kenn03}.
In the spectroscopic observations of the SN
obtained on 2008 September 30 at Calar Alto, light from two H\,{\sc ii}
regions, ``A'' and ``B,'' approximately to the south of
the SN also fell within the CAFOS slit. 

\citet{pet04} calibrate relations between the O abundance and the strong-intensity
line indices $N_2=\log$ [N\,{\sc ii}] $\lambda$6584/H$\alpha$ 
\citep[see, e.g.,][]{ter91,sto94,den02} and
$O3N2=\log$ (([O\,{\sc iii}] $\lambda$5007/H$\beta$)/([N\,{\sc ii}] $\lambda$6584/H$\alpha$)).
These two indices are relatively unaffected by extinction, since they depend on 
ratios of intensities for two spectral lines essentially contiguous in wavelength.
The $1 \sigma$ uncertainties in the O abundance from these two indices are
$\pm 0.18$ and $\pm 0.14$ dex, respectively \citep{pet04}.
We measured the intensities for these various emission lines, using standard routines in
IRAF, and estimate these line indices for these two H\,{\sc ii} regions. 

The resulting values of the indices and the computed abundances are given in 
Table~\ref{hiiregions}. We estimate that our measurement errors lead to uncertainties 
in the abundances of, conservatively, $\pm 0.05$ dex. 
Additionally, we can consider the measurements for the H\,{\sc ii} regions
\#27 from \citet{bibby10} and W13 from \citet{mccall85}. 
These two regions are labeled in Figure~\ref{fighiiregion} and listed in
Table~\ref{hiiregions}. We also include in the table 
the deprojected angular distance of each H\,{\sc ii} region from the SN, 
assuming the values of the host-galaxy inclination and position angle from
\citet{bibby10} and references therein. The linear distances in the table are
assuming a distance to the host galaxy of 3.4 Mpc (see \S~\ref{distance}).

One can see that the O abundances of these H\,{\sc ii} regions are all consistent
with an approximate range of 8.4 to 8.5 dex. The region nearest to ($\sim 200$ pc from) 
the SN, in particular, has 12 + log(O/H) = 8.45.
Assuming the solar photospheric abundance is 12 + log(O/H) $= 8.66 \pm 0.05$
\citep{asp04}, we can infer that the metallicity in the SN environment is  
$\approx 0.6$ solar. The mean O abundance of H\,{\sc ii} regions in the LMC is
12 + log(O/H)  $= 8.37 \pm 0.22$ \citep{russell90}, or $\sim 0.5$ solar. Therefore, the metallicity
at the SN site is most likely not solar, but also is likely not quite as low as the mean LMC nebular
metallicity. However, with the level of uncertainties in the \citet{pet04} calibration relations, in 
our measurements, and in the assumed solar and LMC O abundances,
we analyze the progenitor photometry considering a range of metallicities. 
We will show that the inferred intrinsic properties for the progenitor, nonetheless, 
are most consistent with subsolar metallicity in the SN environment.

\subsection{Initial Mass of the Progenitor}\label{mass}

We have assumed a relatively low extinction toward SN 2008bk, based on
the comparison with SN 1999br (\S \ref{photom}).  We further assume
that this extinction (\S \ref{extinction}) is also appropriate to
apply to the progenitor star. From the Gemini optical data, this star, therefore, 
had an absolute magnitude $M^0_V = -4.94$ and a dereddened color
$(V-I)_0 \approx 2.19$ mag, which imply that the progenitor was a
lower-luminosity (class Ib), early M-type RSG \citep[e.g.,][]{hum84}.

\begin{figure}
\includegraphics[angle=0,scale=0.70]{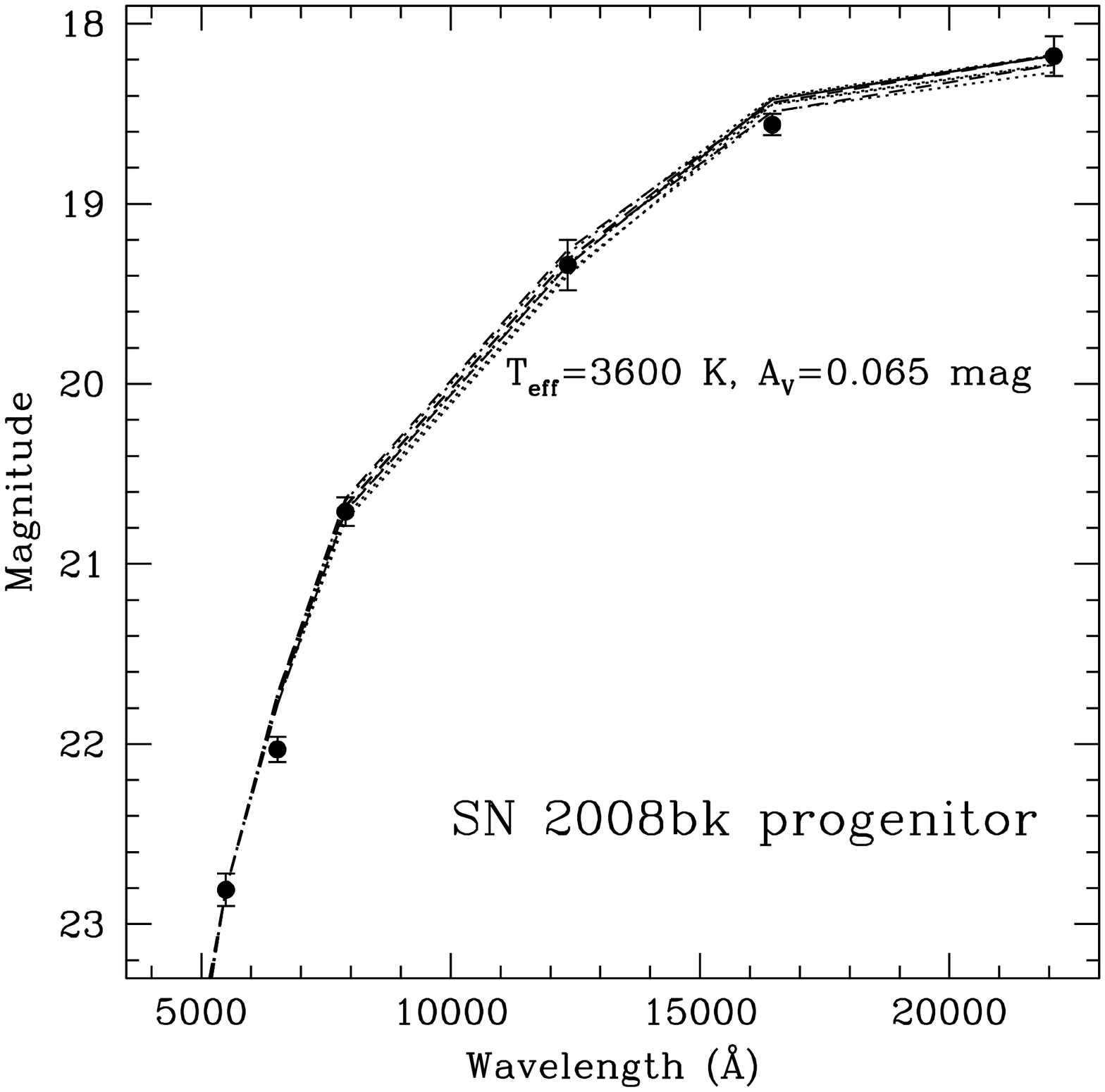}
\caption{The observed spectral energy distribution (SED) of the SN 2008bk
  progenitor star ({\it filled circles}).  For comparison, we also
  show SEDs based on IRAF/SYNPHOT synthetic photometry extracted from
  the MARCS model stellar atmospheres \citep{gus08} at $T_{\rm eff}=3600$ K 
  of 5 and 15 M$_{\odot}$ stars ($\log g=0.0$) at solar metallicity
  ([M/H] = 0)  with microturbulence velocities of 1--5 km s$^{-1}$ 
  ({\it dashed lines}), and at subsolar metallicity ([M/H] = $-0.25$) with 
  microturbulence velocities of 2 and 5 km s$^{-1}$ ({\it dotted lines}). We have reddened the
  model SEDs by our assumed value toward the SN (\S~\ref{extinction}). The models
  were also normalized to the observed brightness of the progenitor in
  $V$. \label{figsed}}
\end{figure}

From the observed optical and near-IR photometry in
Table~\ref{tabprog}, we show the SED of the star in
Figure~\ref{figsed}.  For comparison we show synthetic SEDs, generated
using SYNPHOT applied to the
MARCS\footnote{http://marcs.astro.uu.se/.} model stellar atmospheres
\citep{gus08} for RSGs with effective temperature $T_{\rm eff}=3600$ K and 
surface gravity $\log g=0.0$, assuming spherical geometry
and subsolar metallicity ([M/H] = $-0.25$, where [M/H] denotes the ratio of the
general metal abundance to H, relative to the Sun).
We also show the models at solar metallicity ([M/H] = 0) for this temperature, 
and one can see little difference with the subsolar models.
The MARCS models are for stars with initial masses 5 and 15 M$_{\odot}$ (we will show
that the initial mass of the SN 2008bk progenitor was in between these two values) and have
microturbulence velocities spanning from 1 to 5 km s$^{-1}$.
The models were all reddened by the assumed value for the progenitor, following 
the \citet*{car89} reddening law, and were normalized to the observed $V$ magnitude.
The agreement of all of these models with the observed SED is quite good,
to within the uncertainties in the photometry of the progenitor. (However, the models do diverge
by $\gtrsim\ 1\sigma$ from the observation at $R$, and compare less well with the 
observations at $H$.)  We therefore consider the effective temperature of the star
to be $T_{\rm eff}=3600$ K, with a conservative uncertainty of $\pm 50$ K.

This effective temperature corresponds to spectral type M2--M2.5 
at LMC metallicity \citep[][their Table 4]{levesque06}, and M3 at
solar metallicity \citep[][their Table 5]{levesque05}. 
The metallicity in the SN 2008bk environment is likely somewhere in between (\S~\ref{metals}).
From
\citet[][their Table 6]{levesque05} for Galactic (solar metallicity) RSGs, 
the $V$ and $K$ bolometric corrections at $T_{\rm   eff}=3600$ K are 
${\rm BC}_V=-1.75$ and ${\rm BC}_K=2.84$ mag, respectively.
At LMC metallicity, the bolometric corrections vary little from the
solar values at this $T_{\rm eff}$: ${\rm BC}_V=-1.78$ and ${\rm BC}_K=2.84$ mag
\citep{levesque06}.
The absolute bolometric magnitude of the 
star, therefore, from $V$ is $M_{\rm bol} = -6.70 \pm 0.12$ mag, and from $K$ it is $M_{\rm bol} =
-6.67 \pm 0.13$ mag.  The total uncertainties arise from the uncertainties
in the measured photometry for the star and in the distance modulus (see \S
\ref{distance}), and at $V$, in the range in $BC_V$. That these two estimates of $M_{\rm bol}$,
established from the two different photometric measurements, agree so
well to within the uncertainties, instills us with considerable
confidence in the star's photometry and in our estimates of the
extinction and the star's effective temperature.  
Assuming M$_{\rm bol} (\sun)$ = 4.74 mag, the bolometric
luminosity relative to the Sun is then 
$\log (L_{\rm bol}/{\rm L}_{\odot})=4.57 \pm 0.06$ from both $V$ and $K$.
The luminosity and effective temperature together indicate that the RSG's
radius, at $\sim 6$ months prior to explosion, is $R_{\star} = 496 \pm 34\ {\rm R}_{\odot}$.

The inferred intrinsic properties of the progenitor of SN
2008bk are shown in the HRD in Figure~\ref{fighrd}.
We can also compare the star's locus in the HRD with massive-star
evolutionary tracks of various initial masses, and, given the uncertainties mentioned in 
\S~\ref{metals}, we take into consideration tracks at several metallicities.
In panel (a) of the figure, we compare to tracks at solar metallicity ($Z=0.02$), i.e.,
the Cambridge STARS tracks from \citet{eld04} and the Geneva tracks from \citet*{hirschi04},
which include either stellar equatorial rotation, at $v_{\rm ini} = 300$ km s$^{-1}$, or no rotation.
One can see that these tracks would imply that the initial mass of the progenitor is in the
range of $\sim 10$--12 M$_{\sun}$. However, these tracks all terminate at cooler effective 
temperatures and higher luminosities (and larger stellar radii) than is the case for the progenitor. 
In total, the solar metallicity tracks compare rather poorly with the progenitor's inferred
properties, and we consider it unlikely that the larger initial masses implied by these
tracks apply to the SN 2008bk progenitor.

In Figure~\ref{fighrd}, panel (b), we show a comparison with the STARS tracks at $Z=0.01$
(i.e., at $Z/{\rm Z}_{\sun}=0.5$; the Geneva tracks have not been published
at subsolar metallicity in this mass range). 
The 8 and 8.5 M$_{\sun}$ model tracks 
both terminate with $T_{\rm eff}=3567$ K, 
and with $\log (L_{\rm bol}/{\rm L}_{\odot})=4.59$ and 4.63 (respectively) and 
$R_{\star}=514$ and $541\ {\rm R}_{\odot}$ (respectively), which are essentially 
consistent, to within the uncertainties, with those properties of the progenitor. 
Note that the 9 M$_{\sun}$ model terminates at a luminosity that is higher than
the range we have inferred for the progenitor's luminosity.
Note also that the 7.5 M$_{\sun}$ track terminates at a
much cooler temperature and much higher luminosity, and therefore we can discount it entirely.
In panel (c), we compare with the STARS models for 
$Z=0.008$ (i.e., $Z/{\rm Z}_{\sun}=0.4$), although it is less likely
that the metallicity in the SN environment is quite this low (\S~\ref{metals}). 
The 8 and 8.5 M$_{\sun}$ tracks 
terminate with $T_{\rm eff}=3615$ and 3622 K, $\log (L_{\rm bol}/{\rm L}_{\odot})=4.60$ and 4.63, 
and $R_{\star}=508$ and $527\ {\rm R}_{\odot}$ (respectively), which again agree, to within the
uncertainties, with the star's locus in the HRD (the exception being 
the 7.5 M$_{\sun}$ track, which terminates at a stage 
too cool and too luminous than what is observed, and again, we can discount it).
Much better agreement exists between the star's properties and the subsolar metallicity 
tracks, which, all told, imply that the star's initial mass was $\sim 8$--8.5 M$_{\sun}$. These
stellar evolutionary models all have lost $\sim 0.3\ {\rm M}_{\sun}$ of mass prior to reaching their 
termini.

\subsection{Comparison with \citet{mat08}}

\begin{figure}
\figurenum{10}
\includegraphics[angle=0,scale=0.70]{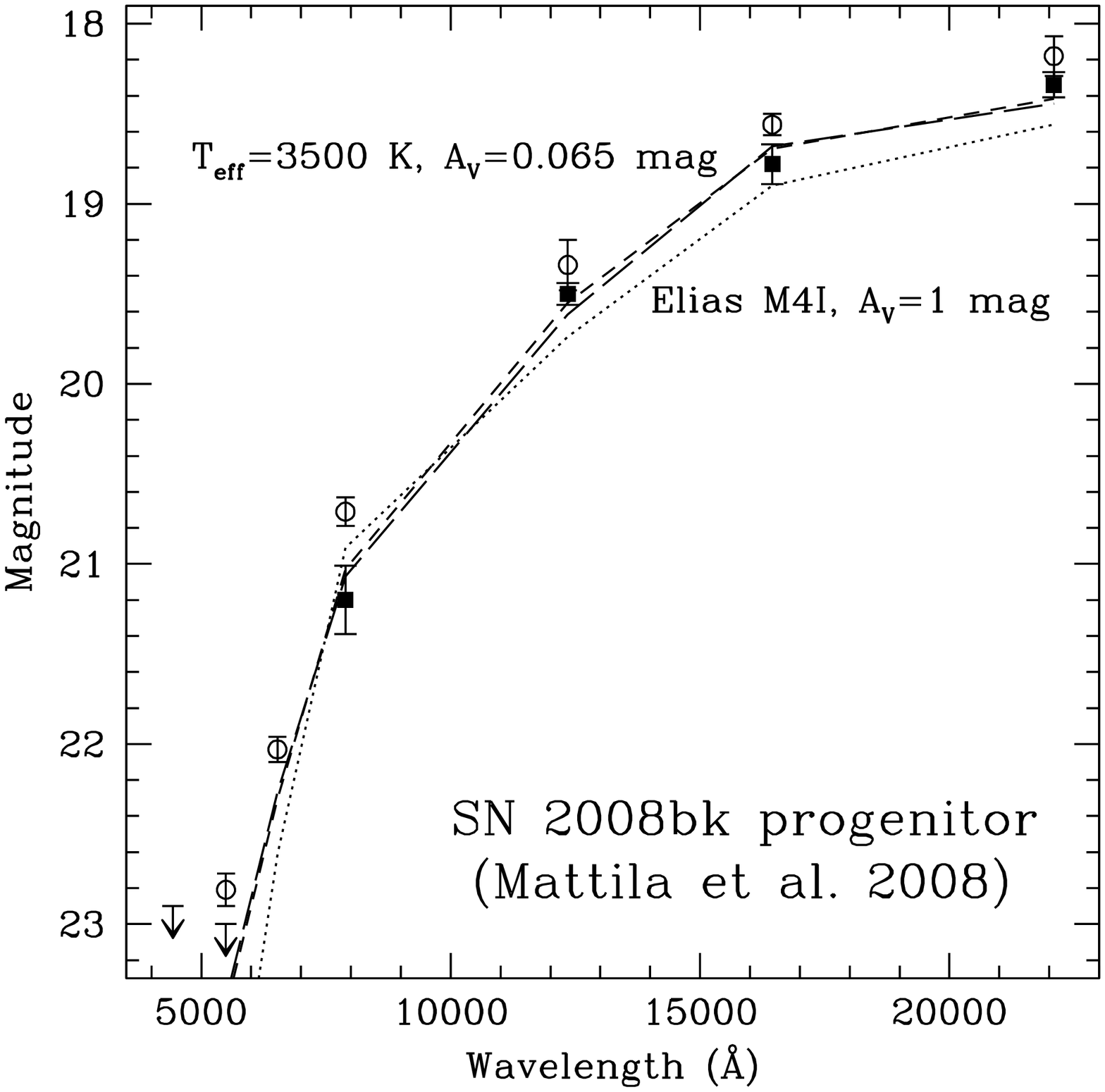}
\caption{The observed SED of the SN 2008bk progenitor star, as measured by  
\citet[][{\it filled squares\/} and {\it upper limits}]{mat08}. We show the SED we have
measured, from Figure~\ref{figsed} ({\it open circles}), for comparison. We also
display the template SED for a M4I supergiant in the LMC 
(\citeauthor{eli85} \citeyear{eli85}; {\it dotted line}), 
extinguished by $A_V=1$ mag \citep{mat08}.
Additionally, we illustrate a synthetic SED derived from MARCS model RSG
stellar atmospheres \citep{gus08} 
with $T_{\rm eff}=3500$ K at subsolar metallicity ([M/H] = $-0.25$), and with 
microturbulence velocities of 2 and 5 km s$^{-1}$ ({\it dashed lines}), 
extinguished by our assumed value toward the SN (\S~\ref{extinction}). \label{figsedmatt}}
\end{figure}

We can make a direct comparison of our results for the SN 2008bk progenitor
with those of \citet{mat08}.
One can see from Figure~\ref{figsedmatt} that all of our magnitude measurements for the star are 
$\gtrsim 1\sigma$ brighter than those measured by \citet{mat08}, particularly their 
measurement at $I$ and their upper limit at $V$ (which is substantially fainter than our 
actual detection in this band). We cannot provide an
explanation for this difference. However, we feel confident in the veracity and self-consistency
of our photometry, based on comparison with pre-existing photometry of stars in the SN
field (\S~\ref{photom}), and on the relatively low dispersions in the transformation of 
$g'r'i'$ photometry to $VRI$ from the Gemini images and of the near-IR instrumental 
photometry to 2MASS magnitudes from the VLT images (\S~\ref{detection}).
Additionally, the \citet{eli85} RSG intrinsic colors, specifically $(V-K)_0$, are systematically bluer,  
for a given $T_{\rm eff}$ and spectral type, than those found by \citet{levesque06}. 
The net 
effect, as seen in Figure~\ref{figsedmatt}, is that higher extinction and reddening are
required to allow for a comparison (and not a particularly good one!) of the model SED for a 
M4I supergiant from \citet{eli85} with
the photometry from \citet{mat08}. Simply for the sake of argument, we also show in the figure the
SED for a RSG from the MARCS stellar atmosphere models with
$T_{\rm eff}=3500$~K (which \citeauthor{mat08}~assume to be the equivalent of a M4I star
at LMC metallicity), adopting our extinction to the SN. This model actually provides 
a better comparison with the \citet{mat08} photometry than do the \citet{eli85} intrinsic colors, 
and this not only would imply that the effective temperature is cooler than what we have found, 
but does so in agreement with a far lower value for the extinction than what \citet{mat08} 
had assumed. The implied cooler temperature is driven primarily by the significantly fainter 
$I$-band magnitude and upper limit to the $V$ brightness found by \cite{mat08}.

\begin{figure}
\figurenum{11}
\includegraphics[angle=0,scale=0.70]{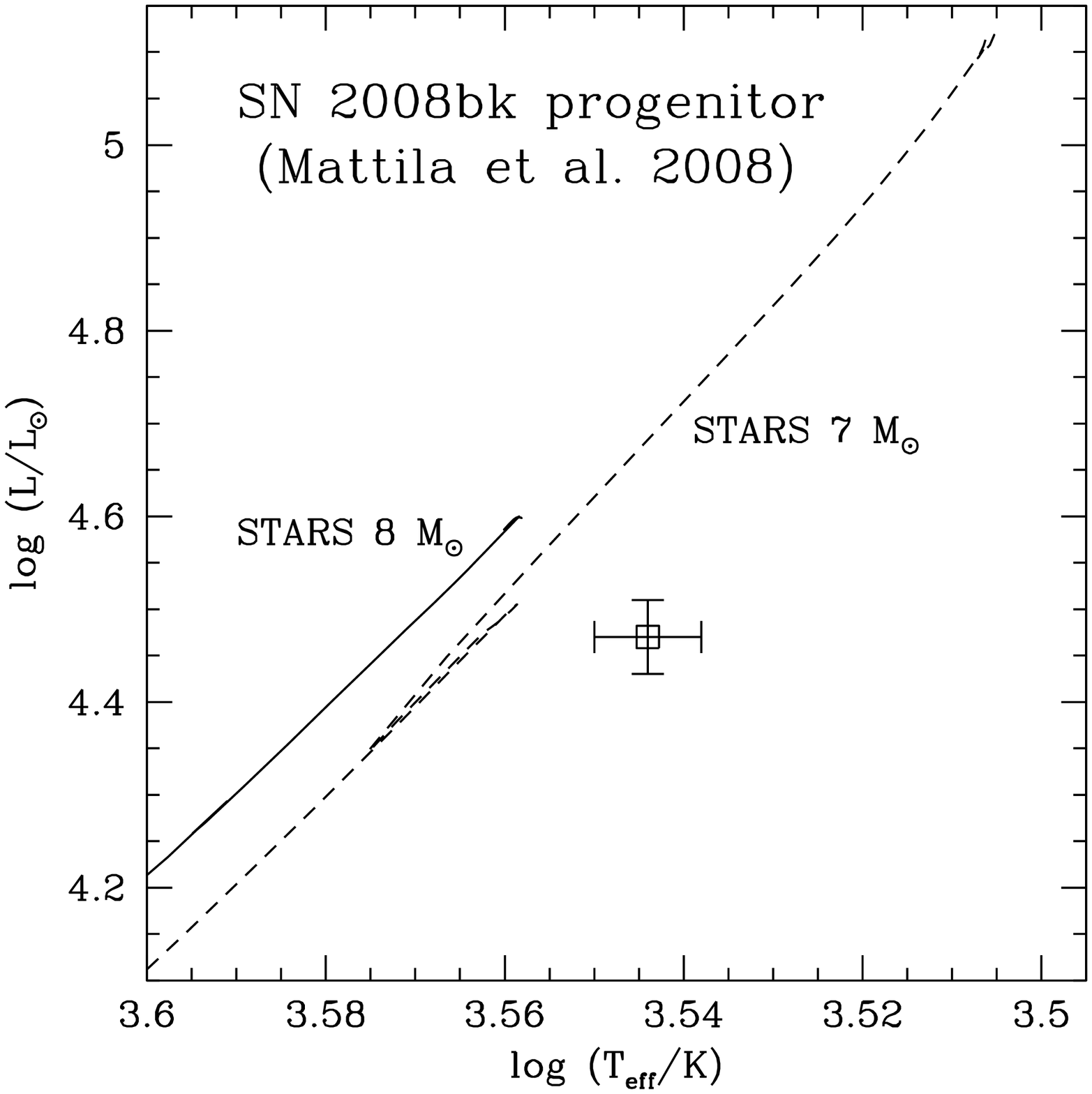}
\caption{An HRD showing the locus of the SN 2008bk progenitor,
based on the $K$-band measurement by \citet{mat08}, and assuming the overall SED
can be modeled by a MARCS model RSG stellar atmosphere 
\citep{gus08} with $T_{\rm eff}=3500$ K at subsolar metallicity ([M/H] = $-0.25$), 
and further assuming our value of extinction toward the SN (see Figure~\ref{figsedmatt}).
For comparison, we also show the STARS massive-star evolutionary tracks \citep{eld04}, 
assuming a subsolar metallicity $Z=0.008$ as \citet{mat08} have, 
at initial masses 7 and 8 M$_{\sun}$ ({\it dashed line} and {\it solid line}, respectively).
\label{fighrdmatt}}
\end{figure}

The fainter $K$-band measurement by \citet{mat08}, combined with the implied cooler 
effective temperature, which corresponds to a larger (positive) $BC_K$ 
\citep{levesque05,levesque06}, as well as the lower value for extinction, all 
lead to a much lower luminosity for the SN progenitor than what we have found to be the case.
A comparison of the luminosity and effective temperature with massive-star evolutionary models in
a HRD shown in Figure~\ref{fighrdmatt} implies that the SN progenitor would 
have an initial
mass well below the threshold mass necessary for core collapse. It is only by virtue of their
assumption that $A_V \approx 1.0$ mag, which, based on their photometry, results in the star 
being more luminous than what we find here, that \citeauthor{mat08}~arrive at its initial mass
being $M_{\rm ini} \approx 8.5\ {\rm M}_{\sun}$. As we presented in \S~\ref{extinction}, we have 
found no evidence for such a large extinction to the SN.

Estimating the reddening to the progenitor through comparison with assumed
intrinsic properties for RSGs is almost certainly fraught with uncertainties. 
\citet{mat08} also attempted to estimate the extinction based on the Balmer decrement for
an H\,{\sc ii} region \citep[W13;][]{mccall85}, which is $1{\farcm}5$ from the SN 2008bk site. 
In general, it is far more accurate to estimate the reddening from the SN
light curves and colors, particularly during the plateau phase, where the colors of SNe II-P
are remarkably similar \citep[e.g.,][]{bersten09}. Although our light curves for SN 2008bk are
very limited in coverage, we therefore believe that the comparison with the more extensive 
SN 1999br curves provide a reliable estimate of the extinction to SN 2008bk.

\bibpunct[, ]{(}{)}{;}{a}{}{;}
We note that \citet{mat08} assumed the distance modulus from \citet{kar03};
however, this contributes relatively little to the overall difference between our two studies.
Furthermore, our assessments of the metallicity in the SN 2008bk environment differ as well.
\citet{mat08} estimated the metallicity,  
assuming the \citet{pilyugin04} metallicity-radial distance relation for
NGC 7793, and concluded that the O abundance was 
12 + log(O/H) $= 8.2 \pm 0.1$.  \citet{bibby10} also have determined the metallicity
gradient in NGC 7793. If we assume the deprojected radial distance, $r$, 
of SN 2008bk given by \citet[][$3{\farcm}47$]{mat08} and the radial distance at 
$B=25$ mag arcsec$^{-2}$, $R_{25}=4{\farcm}65$ \citep[][and references therein]{bibby10},
then for $r/R_{25}=0.75$, the abundance is a slightly higher 12 + log(O/H) $= 8.34 \pm 0.05$.
We have shown in \S~\ref{metals} from direct estimation of the O abundance in 
neighboring H\,{\sc ii} regions that 12 + log(O/H) $\approx 8.4$--8.5.

\section{Discussion and Conclusions}\label{conclusions}

\bibpunct[; ]{(}{)}{;}{a}{}{;}
We conclude that SN 2008bk in NGC 7793 is most likely a
low-luminosity SN~II-P, similar in characteristics to SN 1999br. We
have also confirmed the identification of an RSG as the SN progenitor. 
This star is readily detected in pre-SN, high-quality optical
images from the Gemini-South Observatory. Although SN~1987A is closer, its association
with the star Sk $-$69\arcdeg 202 certain, its multi-band brightness and color known
\citep{isserstedt75}, and, more importantly, its
spectral type unambiguously defined \citep{rousseau78}, 
SN 2008bk ranks as the ``second best-characterized'' SN
progenitor identification to date of any SN type. For this reason, its observed properties are
well constrained.

We have assumed, based on the photometric comparison with SN 1999br,
that the extinction toward SN 2008bk is quite low, $A_V=0.065$ mag,
entirely consistent with Galactic foreground reddening. We have also
estimated the metallicity in the SN environment from neighboring
H\,{\sc ii} regions and find that it is consistent with subsolar
metallicity ($Z \approx 0.6 {\rm Z}_{\sun}$). Furthermore, based on comparison of
the observed SED with recent models for RSG stellar atmospheres, 
we are able to assign an effective temperature of $T_{\rm eff}=3600 \pm 50$ K, and an
absolute bolometric luminosity (relative to the Sun) of $\log (L_{\rm
  bol}/{\rm L}_{\odot})=4.57 \pm 0.06$, both quantities with rather small
uncertainties. We note the very good agreement in the bolometric
magnitudes derived  independently from both the observed $V$ and $K$. 
Through comparison with recent theoretical massive-star 
evolutionary models for subsolar metallicity, we conclude that the progenitor of SN 2008bk had
an initial mass in the range of 8--8.5 M$_{\sun}$.  
We note that this
inferred mass range is consistent with the
upper limit derived for the initial mass of the progenitor of SN
1999br \citep[$< 15\ {\rm M}_{\odot}$;][]{sma09}, and is also within
the initial mass range for RSG progenitors of SNe~II-P, 
$M_{\rm ini}=8.5^{+1}_{-1.5}$ to $16.5 \pm 1.5\ {\rm M}_{\sun}$ \citep{sma09}. 

Further constraints on the initial mass of the SN 2008bk progenitor can be obtained
through comparison with recent theoretical, radiative-transfer/hydrodynamical simulations of 
SNe II-P.
We can compare our estimate of the star's radius near explosion, 
$R_{\star} = 496 \pm 34\ {\rm R}_{\odot}$,  to those input to the pre-SN
models generated {\it at solar metallicity\/} by \citet{dessart10b}: 
their nonrotating $11\ {\rm M}_{\odot}$ model has $R_{\star} = 587\ {\rm R}_{\odot}$, while their rotating
$10\ {\rm M}_{\odot}$ has $R_{\star} = 552\ {\rm R}_{\odot}$. This implies that the SN 2008bk progenitor
had to be $\lesssim 10\ {\rm M}_{\odot}$, consistent with what we infer from comparison with 
theoretical massive-star evolutionary tracks.

The low tail-phase luminosity for SNe~II of the SN 1999br variety has
been explained by a low $^{56}$Ni mass produced in the explosion of a
lower mass star. As \citet{sma09} point out, nuclear statistical
equilibrium is therefore reached only in a thin O-Si-rich layer around
the core in such stars.  Low progenitor masses are consistent with
what has been directly found for other low-luminosity SNe~II-P, such
as SN 1999br, SN 2003gd in M74 \citep{van03,sma04,hen05}, and SN 2005cs
\citep{li06,pas09}.  The lower initial mass estimate for the
progenitor of SN 2008bk presented here 
is consistent with this trend and is essentially
equivalent to the minimum mass for a SN~II-P progenitor \citep[see above;][]{sma09}.  
It is notable that the case of SN 2008bk offers a possible indication that metallicity 
could also play a role in the low luminosity. \citet{sma09} assume in the cases
of SNe 1999br and 2003gd that the metallicity in these environments was subsolar;
on the other hand, they adopt essentially solar metallicity in the case of SN 2005cs.
It will be interesting to see in future cases whether a two-parameter (mass and metallicity) trend 
is possibly emerging.

\begin{figure}
\includegraphics[angle=0,scale=0.70]{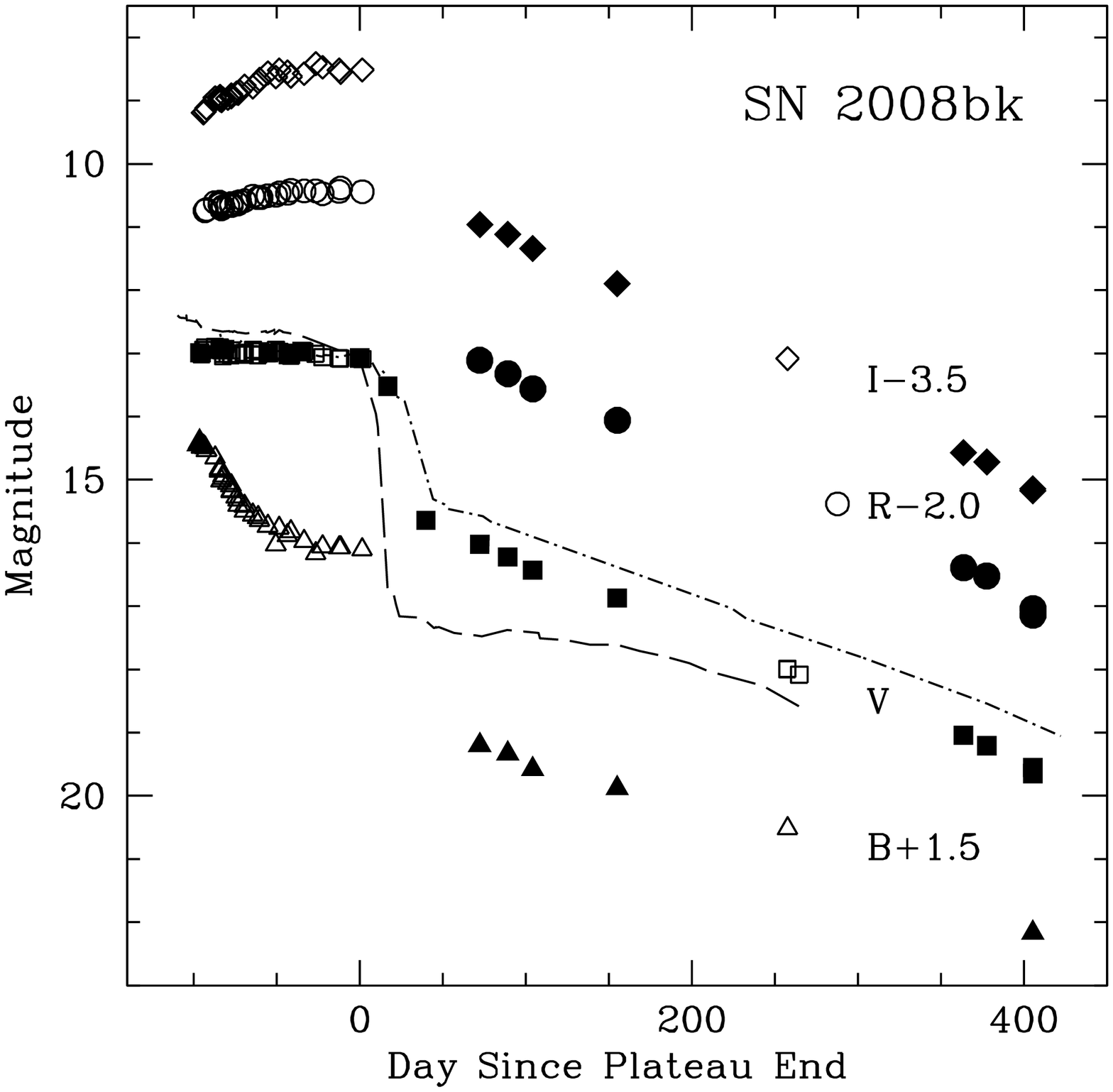}
\caption{The limited $BVRI$ light curves of SN 2008bk from photometry
  in Table~\ref{tabphot} ({\it solid points}). 
  Shown for
  comparison are the light curves for the subluminous SN II-P 1999br
  in NGC 4900 \citep[{\it open points};][]{pas04}, as well as the $V$
  light curves for the SNe II-P 1999em in NGC 1637 ({\it dot-dashed
    line}; \citealt{hamuy01,leo02,elm03}) and 2005cs in M51 ({\it
    dashed line}; \citealt{pas09}). These latter comparison light
  curves have all been adjusted, by eye, in brightness to match the
  end of the plateau phase for SN 2008bk. All light curves are shown
  relative to the epoch corresponding to the end of the plateau phase.\label{figlc}}
\end{figure}

We have already noted that SN 2008bk is $\sim 1.0$ mag more luminous than SN~1999br
\citep[$M^0_V=-13.8$ mag;][]{pas04}.  This is particularly surprising,
given the close overall agreement between the spectra, light-curve
shapes, and color evolution for both SNe 1999br and 2008bk; one is
tempted to expect their luminosities near maximum brightness to at least be
comparable, if not the same (however, see \citeauthor{pas04} \citeyear{pas04} for an assessment
of the diversity among low-luminosity SNe II-P).  One possible reconciliation of 
the different luminosities could be to assume a distance to
the host galaxy (NGC 4900) of SN~1999br which is different from what has been
adopted previously. \citet{pas04} assumed a ``short'' distance (17.3
Mpc), which is essentially that derived from the Tully-Fisher relation
\citep{tully88}; if one adopts the ``long'' distance (39.5 Mpc) found
by \citet{jon09} using the Expanding Photosphere Method (EPM), then
$M^0_V=-15.6$ mag, which is now $\sim 0.8$ mag more luminous than
SN~2008bk.  Although EPM has its own inherent uncertainties and
pitfalls, the implication is that, assuming SNe~2008bk and 1999br are
very similar events, the actual distance to the SN~1999br host may be
somewhere in between the long and short estimates. (Our by-eye
adjustment of the SN 1999br light curves in Figure~\ref{figlc} to
match those of SN~2008bk would then imply that the distance to NGC
4900 could be $\sim 32$ Mpc.)

\begin{figure}
\includegraphics[angle=0,scale=0.60]{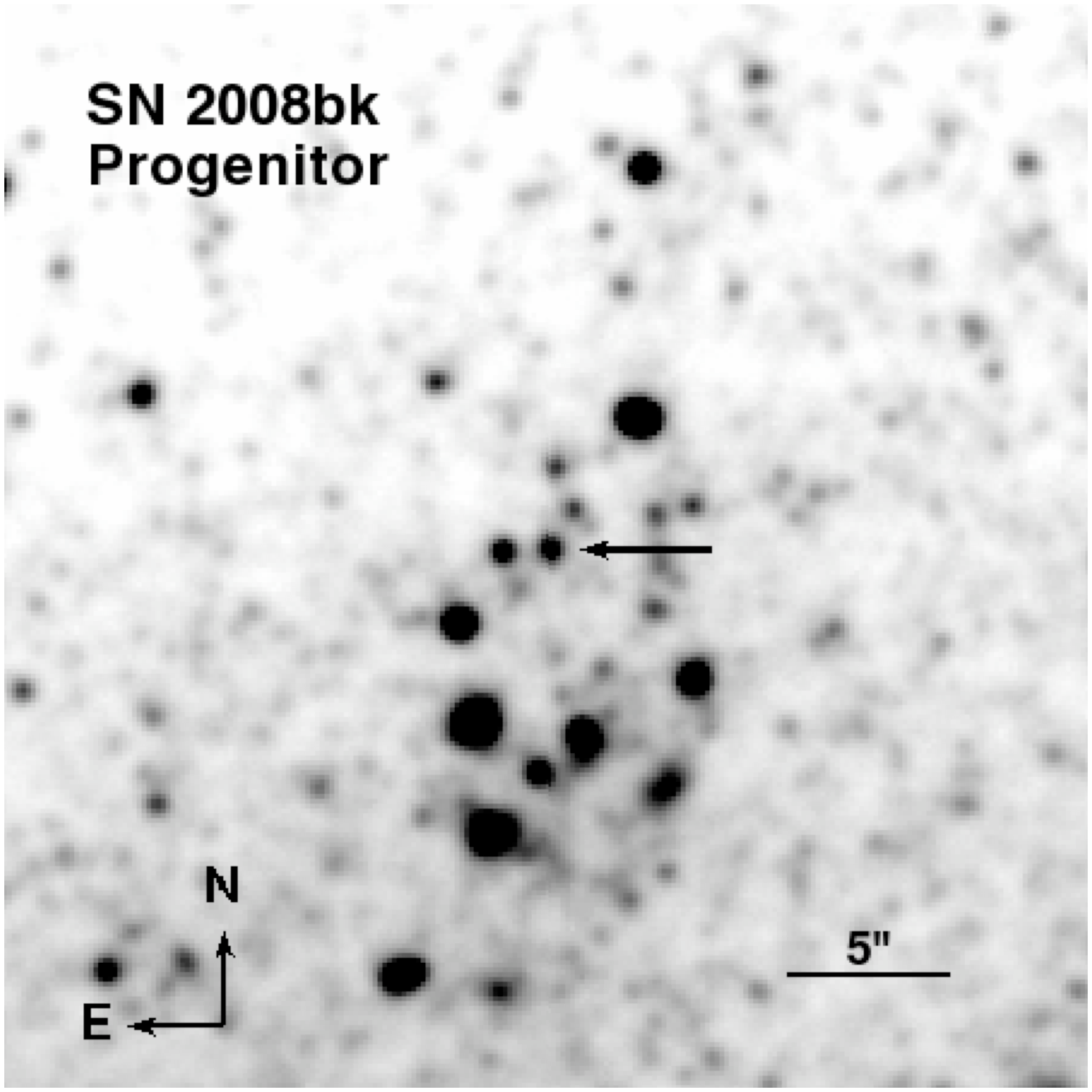}
\caption{The coadded $i'$-band image, from data obtained in 2007
  September and October with the Gemini-South GMOS imager, showing the
  confirmed RSG progenitor of SN 2008bk (indicated by the {\it
    arrow}).  The area of the image displayed here is $33\arcsec \times
  33\arcsec$.  The effective seeing in the combined image is $\sim
  0{\farcs}65$.
\label{figimage}}
\end{figure}

\begin{figure}
\figurenum{12}
\includegraphics[angle=0,scale=0.70]{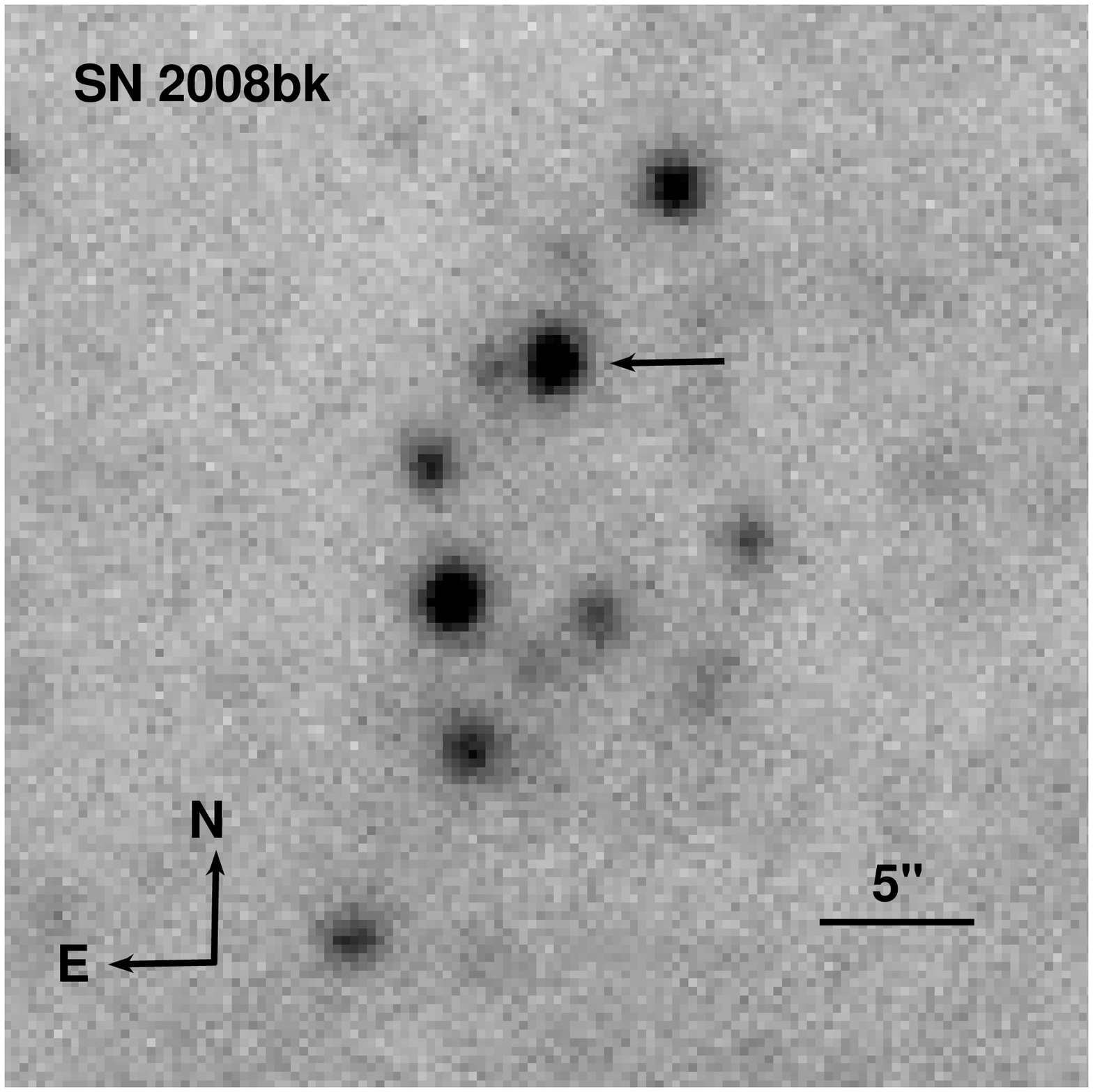}
\caption{The $I$-band image, obtained on 2009 August 13 at the NOT,
  showing SN 2008bk (indicated by the {\it arrow}) at age $\sim 507$~d
  with $I=18.77$ mag (see Table~\ref{tabphot}), only $\sim 2$ mag
  brighter than the progenitor shown in Figure~\ref{figimage}.  The
  scale and orientation here are approximately the same as those in
  the latter figure.  The seeing was $\sim 1{\farcs}0$.
\label{figlater}}
\end{figure}

Finally, although it will be ultimately satisfying to validate the
identification of the progenitor star by obtaining images of SN 2008bk
years hence, when it has faded close to invisibility (e.g., \citeauthor{mau09} 
\citeyear{mau09})\footnote{But note that 
\citet{dessart10a} conjecture that at least 
some low-luminosity SNe~II-P could be SN ``impostors,'' and therefore 
the star could survive what might be an outburst, rather than a 
core-collapse explosion.}, the quality of the pre-SN imaging data and high
astrometric precision of the SN and progenitor positions instill in us
confidence in the star's identification. Furthermore, the
late-time $I$ image from 2009 August (see Table~\ref{tabphot}),
shown in Figure~\ref{figlater}, when compared directly with
Figure~\ref{figimage}, demonstrates with relatively little doubt 
that the progenitor is known. As \citet{mat08} point out, in the
$K_s$-band pre-SN image, once the PSF of the RSG was subtracted away,
a faint source $\sim 0{\farcs}5$ south of the SN position remains (we
also see this same source, when the PSF of the SN is subtracted from
the late-time $I$-band images). Based on the arguments above, we
consider it highly unlikely that this much fainter object is the SN
progenitor.

\acknowledgements

This work is based in part on observations obtained at the Gemini
Observatory, which is operated by the Association of Universities for
Research in Astronomy, Inc., under a cooperative agreement with the
NSF on behalf of the Gemini partnership: the National Science
Foundation (United States), the Science and Technology Facilities
Council (United Kingdom), the National Research Council (Canada),
CONICYT (Chile), the Australian Research Council (Australia),
Ministerio da Ciencia e Tecnologia (Brazil) and Ministerio de Ciencia,
Tecnologia e Innovacion Productiva (Argentina). It is also based in
part on data collected with the 2.2 m telescope of the Calar Alto
Observatory (Sierra de Los Filabres, Spain); on observations made with
the Nordic Optical Telescope, operated on the island of La Palma
jointly by Denmark, Finland, Iceland, Norway, and Sweden, in the
Spanish Observatorio del Roque de los Muchachos of the Instituto de
Astrofisica de Canarias; and on archival data obtained with the 
{\it Spitzer Space Telescope}, which is operated by the Jet Propulsion
Laboratory, California Institute of Technology under a contract with
NASA.

We thank A.~Pastorello for the reductions of the CTIO spectra.  We are
also grateful to S.~Larsen for the photometric calibration of his
Danish telescope images of NGC 7793.  The SN 1999br spectrum was
obtained from the Padova-Asiago Supernova Archive.  This research also
utilized the NASA/IPAC Extragalactic Database (NED), which is operated
by the Jet Propulsion Laboratory, California Institute of Technology,
under contract with NASA. A.V.F. and W.L. are grateful for the support
of NSF grant AST-0908886 and the TABASGO Foundation. 
M.H. acknowledges support by Proyecto Regular Fondecyt 1060808. 
M.H. and G.P acknowledge partial support from Centro de Astrof\'isica FONDAP 15010003,
Center of Excellence in Astrophysics and Associated Technologies PFB 06, and by Programa 
Iniciativa Cient\'ifica Milenio de MIDEPLAN (grant P06-045-F).

\clearpage

\begin{deluxetable}{ccccc}
\tablewidth{4.5truein}
\tablecolumns{5}
\tablecaption{Photometric Sequence around SN 2008bk\tablenotemark{a}\label{tabfids}}
\tablehead{
\colhead{Star} & \colhead{$B$ (mag)} & \colhead{$V$ (mag)} &
\colhead{$R$ (mag)} & \colhead{$I$ (mag)}
}
\startdata
A & 17.459(004) & 16.328(003) & 15.728(004) & 15.085(008) \\
B & 16.939(003) & 16.283(003) & 15.910(003) & 15.447(008) \\
C & 18.894(012) & 18.120(015) & 17.662(015) & 17.132(013) \\
D & 18.574(005) & 17.898(003) & 17.512(004) & 17.045(008) \\
E & 18.184(005) & 17.314(003) & 16.813(003) & 16.290(010) \\
F & 19.415(007) & 18.251(003) & 17.531(003) & 16.832(008) \\
G & 19.025(006) & 17.714(003) & 16.836(003) & 15.861(008) \\
H & 20.092(011) & 18.642(004) & 17.664(000) & 16.561(008) \\
I & 19.392(008) & 18.054(003) & 17.178(003) & 16.219(006) \\
J & 17.512(012) & 17.401(012) & 17.289(012) & 16.918(011) \\
K & 15.013(001) & 14.278(018) & 15.080(051) & 13.497(014) \\
L & 16.794(004) & 16.130(004) & 15.790(004) & 15.364(007) \\
M & 17.515(004) & 17.060(004) & 16.814(004) & 16.444(007) \\
N & 18.730(004) & 18.031(004) & 17.649(004) & 17.165(004) \\
\enddata
\tablenotetext{a}{Uncertainties $(1\sigma)$ are given in 
parentheses, in units of 0.001 mag.}
\end{deluxetable}

%\clearpage

\begin{deluxetable}{lcccccc}
\tablewidth{6.7truein}
\tablecolumns{7}
\tablecaption{Optical Photometry of SN 2008bk\tablenotemark{a}\label{tabphot}}
\tablehead{
\colhead{UT date} &\colhead{JD$-$2450000 } & \colhead{$B$ (mag)} &
\colhead{$V$ (mag)} & \colhead{$R$ (mag)} & \colhead{$I$ (mag)} & 
\colhead{Source\tablenotemark{b}}}
\startdata
2008 Mar 28.41 & 4553.90 & 12.94(13) & \nodata & \nodata & \nodata & CHASE \\
2008 Mar 29.40 & 4554.90 & 12.89(14) & 12.99(10) & \nodata & \nodata & CHASE \\
2008 Apr 13.41 & 4569.91 & \nodata & 12.94(05) & \nodata & \nodata & LCO duPont \\
2008 May 10.79 & 4596.29 & \nodata & 12.99(13) & \nodata & \nodata & S.~Howerton \\
2008 May 22.80 & 4608.30 & \nodata & 13.02(06) & \nodata & \nodata & S.~Howerton \\
2008 May 30.85 & 4616.35 & \nodata & 12.97(26) & \nodata & \nodata & S.~Howerton \\
2008 Jul 4.80 & 4651.30 & \nodata & 13.07(12) & \nodata & \nodata & S.~Howerton \\
2008 Jul 21.65 & 4668.15 & \nodata & 13.52(13) & \nodata & \nodata & S.~Howerton \\
2008 Aug 13.63 & 4691.13 & \nodata & 15.64(11) & \nodata & \nodata & S.~Howerton \\
2008 Sep 13.02 & 4723.52 & 17.70(07) & 16.02(06) & 15.11(05) & 14.46(08) & Calar Alto \\
2008 Sep 30.93 & 4740.43 & 17.83(07) & 16.22(06) & 15.32(05) & 14.62(08) & Calar Alto\\
2008 Oct 15.91 & 4755.41 & 18.08(08) & 16.43(06) & 15.56(05) & 14.84(08) & Calar Alto\\
2008 Dec 5.84 & 4806.34 & 18.38(07) & 16.87(06) & 16.06(05) & 15.40(08) & NOT \\
2009 Jul 02.37 &  5014.87 & \nodata & 19.04(03) & 18.39(03) & 18.07(03) & SMARTS \\
2009 Jul 16.36 &  5028.86 & \nodata & 19.21(05) & 18.52(07) & 18.22(03) & SMARTS \\
2009 Aug 13.13 & 5056.63 & 20.67(06) & 19.55(08) & 19.14(05) & 18.67(08) & Calar Alto \\
2009 Aug 13.16 & 5056.66 & \nodata & 19.65(05) & 19.04(03) & 18.64(03) & NOT \\
\enddata
\tablenotetext{a}{Uncertainties $(1\sigma)$ are given in parentheses, in
units of 0.01 mag.}
\tablenotetext{b}{CHASE = 0.41 m PROMPT2 + Apogee U47; 
LCO duPont = Las Campanas Observatory 2.5 m duPont + SITe2K; 
S.~Howerton = GRAS-008 + SBIG ST-L-11K 3, GRAS-015 + SBIG ST-8 3 or SBIG ST-10 3 CCD; 
Calar Alto = 2.2 m Calar Alto + CAFOS; 
NOT = 2.5 m Nordic Optical Telescope + ALFOSC; SMARTS = 1.3 m SMARTS + ANDICAM.}
\end{deluxetable}

%\clearpage

\begin{deluxetable}{lccccc}
\tablewidth{0pt}
\tablecolumns{6}
\tablecaption{Optical Spectroscopy of SN 2008bk\label{tabspec}}
\tablehead{
\colhead{UT Date} & \colhead{JD$-$2450000} & \colhead{Day\tablenotemark{a}} & \colhead{Grism/Grating} & \colhead{Wavelength}   & \colhead{Instrument\tablenotemark{b}}\\
\colhead{} & \colhead{} & \colhead{} & \colhead{} & \colhead{Range (\AA)}   & \colhead{}}
\startdata
2008 Jul 23.36  & 4670.86 & 20 
& KPGL3-1  & 4005--7738 & CTIO \\
2008 Jul 24.35  & 4671.85 & 21 
& KPGL3-1  & 4005--7738 & CTIO \\
2008 Jul 25.31  & 4672.81 & 22 
& KPGL3-1  & 4005--7738 & CTIO \\
2008 Sep 14.04 & 4723.54 & 72 
& G200 & 4900--10458 & CAHA \\
2008 Sep 16.02 & 4725.52 & 74 
& G200 & 3665--10463 & CAHA \\
2008 Sep 30.93 & 4740.43 & 89 
& R100 & 5710--9585 & CAHA \\
\enddata
\tablenotetext{a}{The approximate day is referenced to our estimate of the end of the plateau phase, JD $\approx$ 2454651.3.}
\tablenotetext{b}{CTIO = Blanco 4 m telescope + R-C Spec 
(Cerro Tololo, Chile); CAHA = Calar Alto 2.2 m telescope + 
CAFOS SiTe (near Almeria, Spain).}
\end{deluxetable}

%\clearpage

\begin{deluxetable}{cc}
\tablewidth{0pt}
\tablecolumns{2}
\tablecaption{Photometry of the SN 2008bk Progenitor\tablenotemark{a}\label{tabprog}}
\tablehead{
\colhead{Band\phantom{$_s$}} & \colhead{Magnitude} }
\startdata
$V$\phantom{$_s$} & 22.81(09) \\
$R$\phantom{$_s$}  & 22.03(07) \\
$I$\phantom{$_s$}  & 20.71(08) \\
$J$\phantom{$_s$}  & 19.26(14) \\
$H$\phantom{$_s$}  & 18.55(06) \\
$K_s$ & 18.14(10) \\
\enddata
\tablenotetext{a}{Uncertainties $(1\sigma)$ 
are given in parentheses, in units of 0.01 mag. 
The $JHK_s$ magnitudes are in the 2MASS photometric system.}
\end{deluxetable}

%\clearpage

\begin{deluxetable}{cccccccc}
%\rotate
\tablewidth{0pt}
\tablecolumns{8}
\tablecaption{Strong-line Indices and Oxygen Abundances\tablenotemark{a} 
for H\,{\sc ii} Regions Near SN 2008bk\label{hiiregions}}
\tablehead{
\colhead{Region} & \colhead{SN offset ($\arcsec$)\tablenotemark{b}} & 
\colhead{SN offset (pc)\tablenotemark{c}} 
& \colhead{\phantom{$<$}$N_2$} & 12 + log(O/H) & \colhead{\phantom{$<$}$O3N2$} & 
\colhead{\phantom{$<$}12 + log(O/H)} & \colhead{Source}}
\startdata
A & 12.3 & 203 & \phantom{$<$}$-0.71$ &  \phantom{$<$}8.45 & 
\phantom{$<$}\nodata & \phantom{$<$}\nodata & This work\\
B & 95.6 & 1576 & \phantom{$<$}$-0.62$ & \phantom{$<$}8.52  & 
\phantom{$<$}\nodata & \phantom{$<$}\nodata & This work\\
\#27 & 82.2 & 1355 & \phantom{$<$}$-0.66$ & \phantom{$<$}8.53 & 
\phantom{$<$}$0.81$ & \phantom{$<$}8.47 & \citet{bibby10} \\
W13 & 103.2 & 1701 & $<$$-0.77$ & $<$8.41 & $>$1.10 & $<$8.38 & \citet{mccall85} \\
\enddata
\tablenotetext{a}{Computed following \citet{pet04}.}
\tablenotetext{b}{Angular distance of the H\,{\sc ii} region from SN 2008bk, deprojected for the host
galaxy's position angle and inclination \citep[adopting the values for these from][]{bibby10}.
See Figure~\ref{fighiiregion}.}
\tablenotetext{c}{Assuming a distance to the host galaxy of 3.4 Mpc \citep{piet10}.}
\end{deluxetable}

\clearpage

\clearpage

\clearpage

\clearpage

\clearpage

\clearpage

\clearpage

\clearpage

\clearpage

\begin{figure}
\figurenum{9}
\includegraphics[angle=0,scale=0.65]{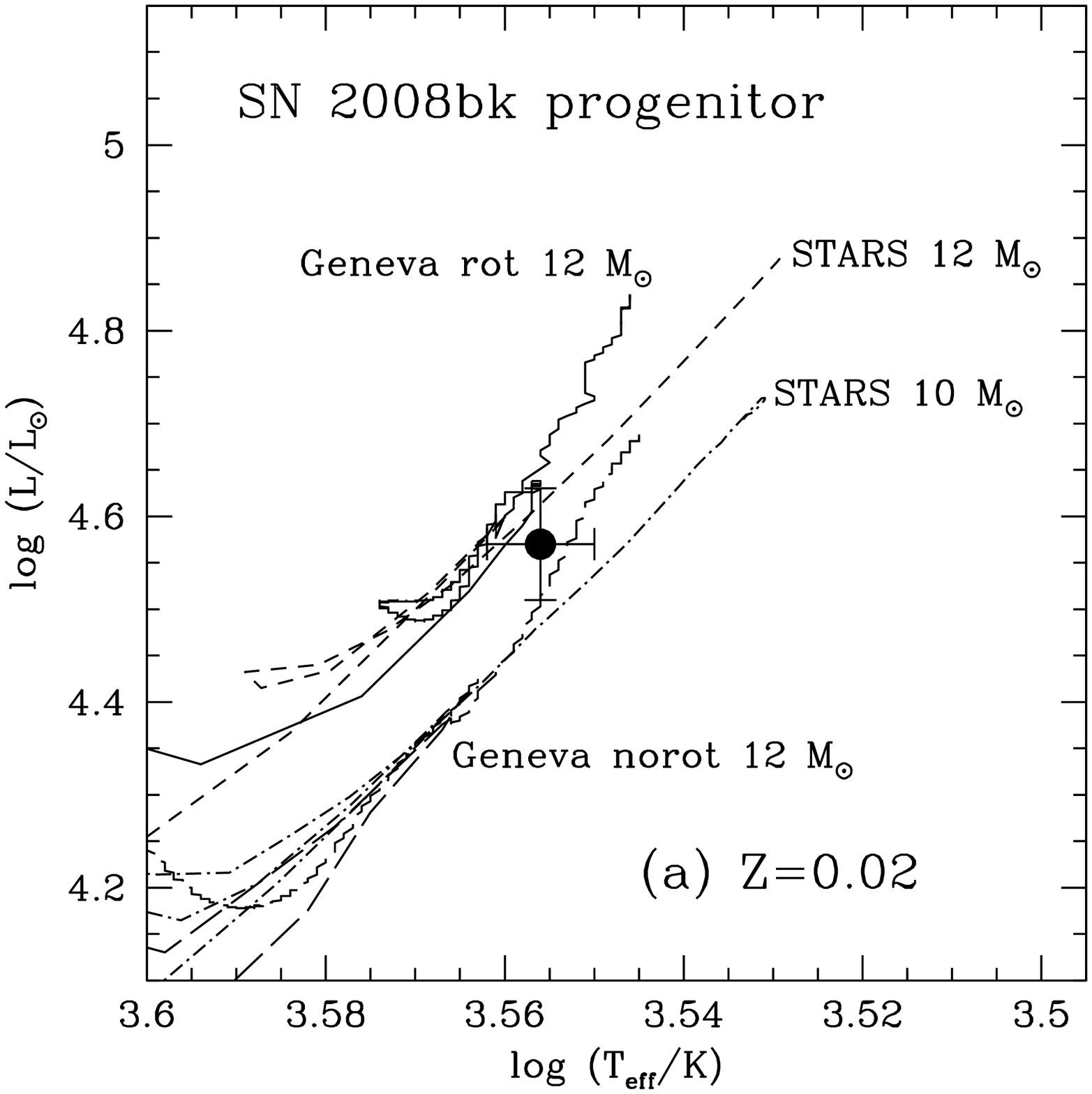}
\caption{An HRD showing the locus of the SN 2008bk progenitor.
  For comparison, we indicate model stellar evolutionary tracks,
  in panel (a) at metallicity $Z = 0.02$, in (b) at metallicity $Z=0.01$, and 
  in (c) at metallicity $Z=0.008$. In (a), the 12 M$_{\odot}$ models are from 
  \citet[][``Geneva'' models]{hirschi04}, 
  with rotation (``rot,'' $v_{\rm ini} = 300$ km s$^{-1}$; {\it solid line}) and 
  without rotation (``norot;'' {\it long-dashed line}), 
  and from \citet[][``STARS'' models]{eld04}, at initial masses 10 M$_{\sun}$ 
  ({\it dot-dashed line}) and 12 M$_{\sun}$ ({\it short-dashed line}). 
  In (b), we show just the STARS models at initial
  masses 7.5, 8, 8.5, and 9 M$_{\sun}$ 
  ({\it dotted, long-dashed, solid, and short-dashed lines}, respectively).
  In (c), we again show just the STARS models at initial masses 7.5, 8, and 8.5 M$_{\sun}$ 
  ({\it dotted, long-dashed, and solid lines}, respectively).
  \label{fighrd}}
\end{figure}

\clearpage

\begin{figure}
\figurenum{9}
\includegraphics[angle=0,scale=0.65]{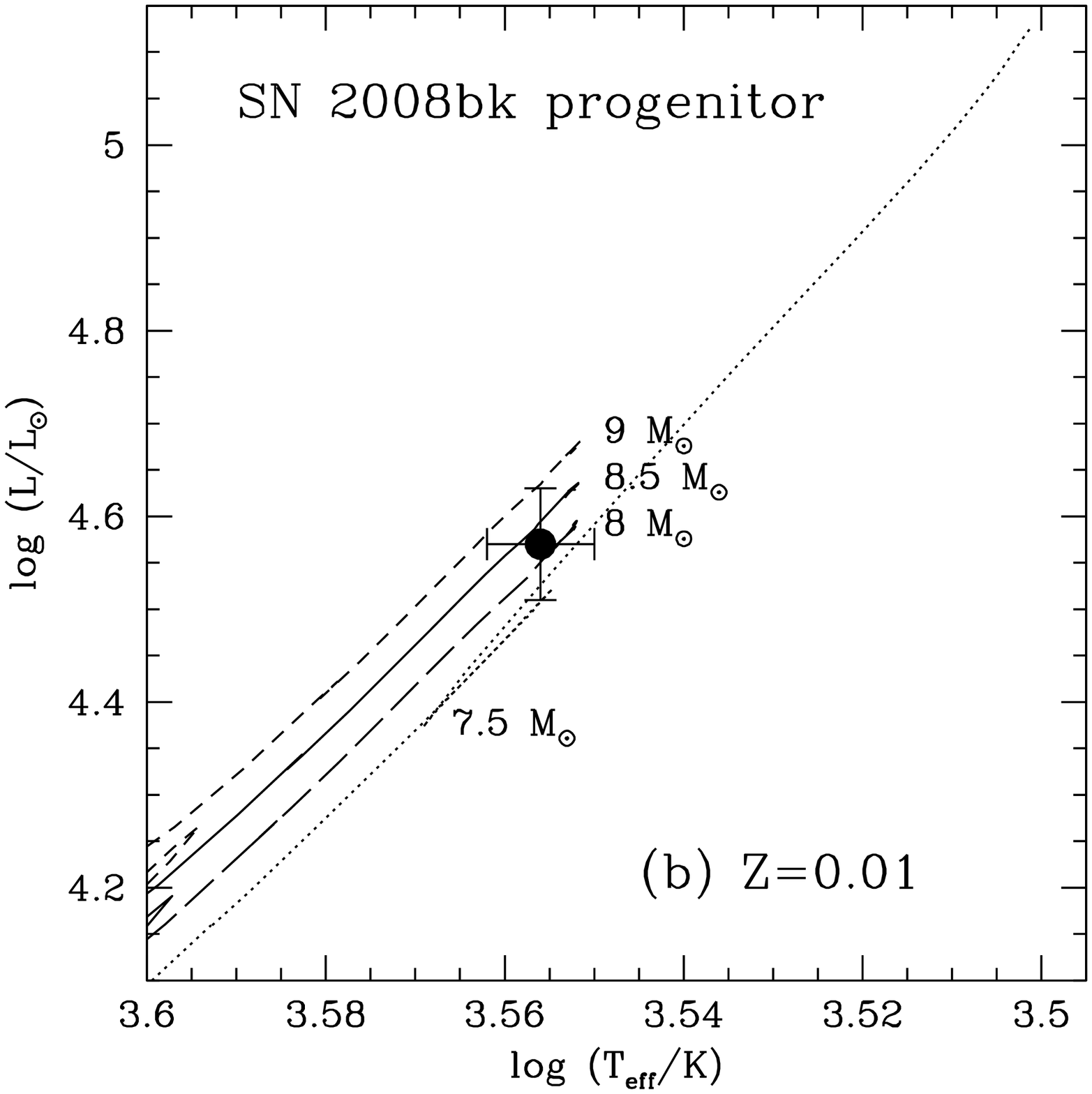}
\caption{(Continued.)\label{fighrdb}}
\end{figure}

\clearpage

\begin{figure}
\figurenum{9}
\includegraphics[angle=0,scale=0.65]{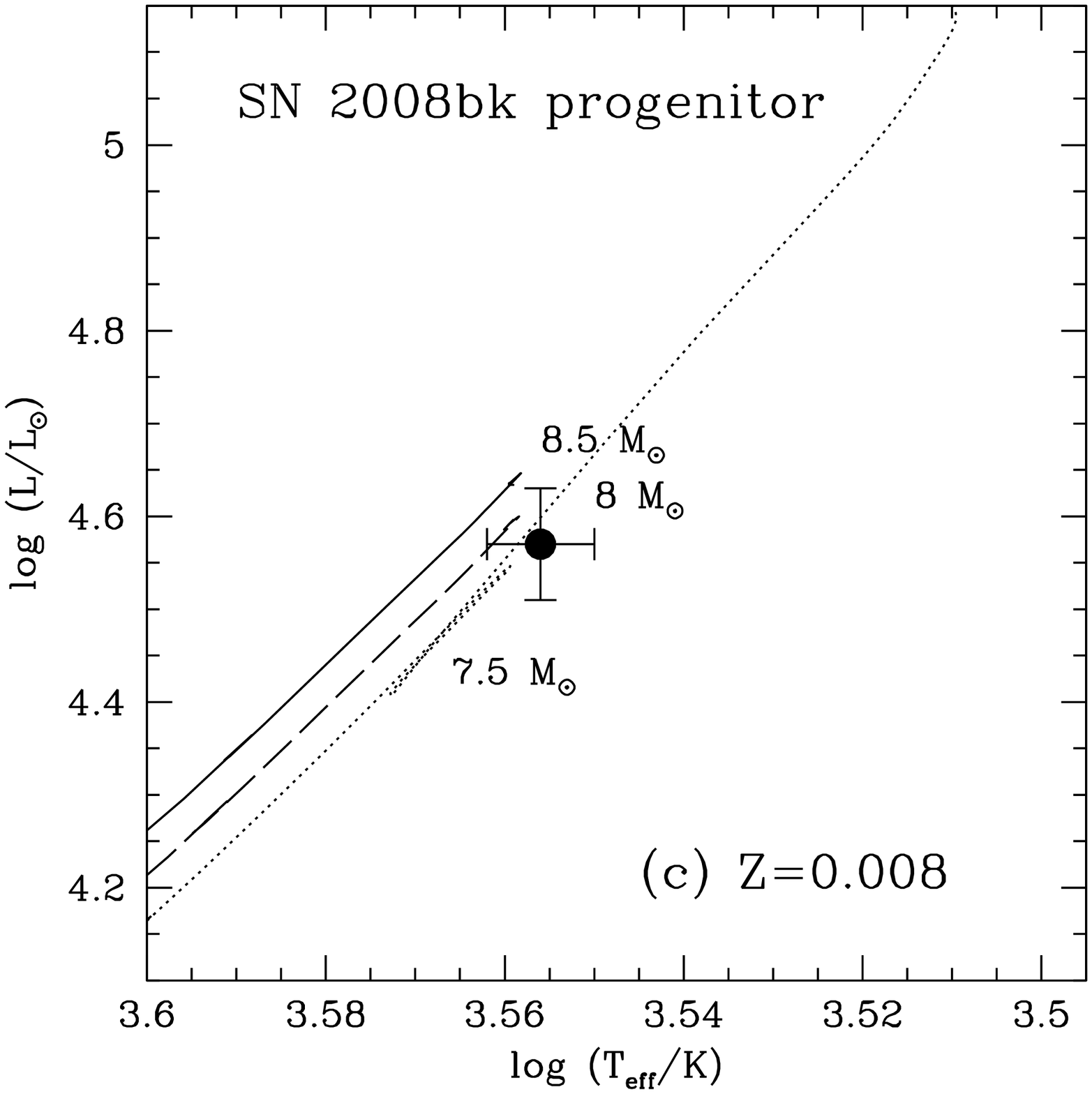}
\caption{(Continued.)\label{fighrdc}}
\end{figure}

\clearpage

\clearpage

\clearpage

\begin{thebibliography}{}

\bibitem[Aldering, Humphreys, \& Richmond(1994)]{ald94} Aldering, G., Humphreys, R. M.,
  \& Richmond, M. W. 1994, \aj, 107, 662
\bibitem[Asplund et al.(2004)]{asp04} Asplund, M., Grevesse, N., Sauval, A.~J., Allende Prieto, C., 
\& Kiselman, D.\ 2004, \aap, 417, 751
\bibitem[Bersten \& Hamuy(2009)]{bersten09} Bersten, M.~C., \& Hamuy, M.\ 2009, \apj, 701, 200 
\bibitem[Bertola(1964)]{ber64} Bertola, F. 1964, AnnAp, 27, 319
\bibitem[Bessell \& Brett(1988)]{bessell88} Bessell, M.~S., \& Brett, J.~M.\ 1988, \pasp, 100, 1134 
\bibitem[Bibby \& Crowther(2010)]{bibby10} Bibby, J.~L., \& Crowther, P.~A.\ 2010, \mnras, 405,
2737
\bibitem[Cardelli, Clayton, \& Mathis(1989)]{car89} Cardelli, J.~A.,
  Clayton, G.~C., \& Mathis, J.~S. 1989, \apj, 345, 245
\bibitem[Carpenter(2001)]{carp01} Carpenter, J.~M.\ 2001, \aj, 121, 2851 
\bibitem[Cohen, Darling, \& Porter(1995)]{coh95} Cohen, J. G.,
  Darling, J., \& Porter, A. 1995, AJ, 110, 308
\bibitem[Crampton et al.(2000)]{cra00} Crampton, D., 
Fletcher, J.~M., Jean, I., et al.\ 2000, \procspie, 4008, 114
\bibitem[Denicol\'o, Terlevich, \& Terlevich(2002)]{den02} Denicol\'o,
  G., Terlevich, R., \& Terlevich, E.  2002, \mnras, 330, 69
\bibitem[DENIS Consortium(2005)]{denis} The DENIS Consortium.\ 2005, VizieR Online Data 
Catalog, 2263, 0
\bibitem[Dessart, Livne, \& Waldman(2010a)]{dessart10a} Dessart, L., Livne, E., 
\& Waldman, R.\ 2010a, \mnras, 405, 2113
\bibitem[Dessart, Livne, \& Waldman(2010b)]{dessart10b} Dessart, L., Livne, E., 
\& Waldman, R.\ 2010b, \mnras, 408, 827 
\bibitem[Eldridge \& Tout(2004)]{eld04} Eldridge, J.~J., \& Tout,
  C.~A. 2004, \mnras, 348, 201
\bibitem[Elias, Frogel, \& Humphreys(1985)]{eli85} Elias, J.~H.,
  Frogel, J.~A., \& Humphreys, R.~M. 1985, \apjs, 57, 91
\bibitem[Elmhamdi et al.(2003)]{elm03} Elmhamdi, A., 
Danziger, I.~J., Chugai, N., et al.\ 2003, \mnras, 338, 939
\bibitem[Fraser et al.(2011)]{fraser11} Fraser, M., Ergon, M., 
Eldridge, J.~J., et al.\ 2011, \mnras, 417, 1417
\bibitem[Gilmozzi et al.(1987)]{gil87} Gilmozzi, R., 
Cassatella, A., Clavel, J., et al.\ 1987, \nat, 328, 318
\bibitem[Gustafsson et al.(2008)]{gus08} Gustafsson, B., Edvardsson,
  B., Eriksson, K., et al.\ 2008,  \aap, 486, 951
\bibitem[Hamuy et al.(2001)]{hamuy01} Hamuy, M., Pinto, P.~A., 
Maza, J., et al.\ 2001, \apj, 558, 615
\bibitem[Heger et al.(2003)]{heger03} Heger, A., Fryer, C.~L., 
Woosley, S.~E., Langer, N., \& Hartmann, D.~H.\ 2003, \apj, 591, 288
\bibitem[Hendry et al.(2005)]{hen05} Hendry, M.~A., Smartt, 
S.~J., Maund, J.~R., et al.\ 2005, \mnras, 359, 906
\bibitem[Hirschi, Meynet, \& Maeder(2004)]{hirschi04} Hirschi, R.,
  Meynet, G., \& Maeder, A. 2004, \aap, 425, 649
\bibitem[Humphreys \& McElroy(1984)]{hum84} Humphreys, R.~M., \&
  McElroy, D.~B. 1984, \apj, 284, 565
\bibitem[Isserstedt(1975)]{isserstedt75} Isserstedt, J.\ 1975, \aaps, 19, 259
\bibitem[Jacobs et al.(2009)]{jacobs09} Jacobs, B.~A., Rizzi, L.,
  Tully, R.~B., Shaya, E.~J., Makarov, D.~I., et al.\ 2009,
  \aj, 138, 332
\bibitem[Jones et al.(2009)]{jon09} Jones, M.~I., Hamuy, M., 
Lira, P., et al.\ 2009, \apj, 696, 1176 
\bibitem[Karachentsev et al.(2003)]{kar03} Karachentsev, I.~D., Grebel, E.~K., Sharina, M.~E., et 
al.\ 2003, \aap, 404, 93
\bibitem[Kennicutt et al.(2003)]{kenn03} Kennicutt, R.~C., 
Jr., Armus, L., Bendo, G., et al.\ 2003, \pasp, 115, 928
\bibitem[Kochanek, Szczygiel, \& Stanek(2011)]{kochanek11} Kochanek, C.~S.,
  Szczygiel, D.~M., \& Stanek, K.~Z. 2011, \apj, 737, 76
\bibitem[Larsen(1999)]{lar99a} Larsen, S. S. 1999, \aaps, 139, 393
\bibitem[Larsen \& Richtler(1999)]{lar99b} Larsen, S. S., \& Richtler,
  T. 1999, \aap, 345, 59
\bibitem[Leonard et al.(2002)]{leo02} Leonard, D.~C., 
Filippenko, A.~V., Gates, E.~L., et al.\ 2002, \pasp, 114, 35
\bibitem[Leonard et al.(2003)]{leo03} Leonard, D.~C., Kanbur, 
S.~M., Ngeow, C.~C., \& Tanvir, N.~R.\ 2003, \apj, 594, 247
\bibitem[Levesque et al.(2005)]{levesque05} Levesque, E.~M., 
Massey, P., Olsen, K.~A.~G., et al.\ 2005, \apj, 628, 973
\bibitem[Levesque et al.(2006)]{levesque06} Levesque, E.~M., 
Massey, P., Olsen, K.~A.~G., et al.\ 2006, \apj, 645, 1102
\bibitem[Li et al.(2006)]{li06} Li, W., Van Dyk, S.~D., 
Filippenko, A.~V., et al.\ 2006, \apj, 641, 1060 
\bibitem[Li et al.(2008)]{li08} Li, W., Van Dyk, S.~D., Filippenko, A.~V., et al.\ 2008, Central Bureau 
Electronic Telegrams, 1319, 1
\bibitem[Li et al.(2011)]{li11} Li, W., Bloom, J.~S., 
Podsiadlowski, P., et al.\ 2011, \nat, in press (arXiv:1109.1593)
\bibitem[Maoz \& Mannucci(2008)]{maoz08} Maoz, D., \& Mannucci, F.\ 2008, The Astronomer's 
Telegram, 1464, 1
\bibitem[Matheson et al.(2000)]{math00} Matheson, T., 
Filippenko, A.~V., Barth, A.~J., et al.\ 2000, \aj, 120, 1487
\bibitem[Mattila et al.(2008)]{mat08} Mattila, S., Smartt, 
S.~J., Eldridge, J.~J., et al.\ 2008, \apjl, 688, L91
\bibitem[Maund \& Smartt(2009)]{mau09} Maund, J.~R., \& Smartt,
  S.~J. 2009, Science, 324, 486
\bibitem[Maund, Smartt, \& Danziger(2005)]{maund05} Maund, J.~R., Smartt, 
S.~J., \& Danziger, I.~J.\ 2005, \mnras, 364, L33 
\bibitem[McCall, Rybski, \& Shields(1985)]{mccall85} McCall, M.~L., Rybski, 
P.~M., \& Shields, G.~A.\ 1985, \apjs, 57, 1
\bibitem[Monard(2008)]{mon08} Monard, L.~A.~G.\ 2008, Central 
Bureau Electronic Telegrams, 1315, 1
\bibitem[Morrell \& Stritzinger(2008)]{mor08} Morrell, N., \& Stritzinger, M.\ 2008, Central Bureau 
Electronic Telegrams, 1335, 1 
\bibitem[Mould \& Sakai(2008)]{mou08} Mould, J., \& Sakai, S.\ 2008,
  \apjl, 686, L75
\bibitem[Nomoto et al.(2006)]{nomoto06} Nomoto, K., Tominaga, 
N., Umeda, H., Kobayashi, C., \& Maeda, K.\ 2006, Nuclear Physics A, 777, 424 
\bibitem[Pastorello et al.(2004)]{pas04} Pastorello, A., 
Zampieri, L., Turatto, M., et al.\ 2004, \mnras, 347, 74
\bibitem[Pastorello et al.(2009)]{pas09} Pastorello, A., 
Valenti, S., Zampieri, L., et al.\ 2009, \mnras, 394, 2266
\bibitem[Pettini \& Pagel(2004)]{pet04} Pettini, M., \& Pagel,
  B.~E.~J.\ 2004, \mnras, 348, L59
\bibitem[Pierce \& Tully(1992)]{pie92} Pierce, M.~J., \& Tully,
  R.~B. 1992, \apj, 387, 47
\bibitem[Pietrzy{\'n}ski et al.(2010)]{piet10} Pietrzy{\'n}ski, G., Gieren, W., Hamuy, M., et al.\ 2010, \aj, 140, 1475
\bibitem[Pignata et al.(2008)]{pig08} Pignata, G., Maza, J., 
Hamuy, M., et al.\ 2008, Central Bureau Electronic Telegrams, 1319, 2
\bibitem[Pilyugin, V{\'{\i}}lchez, \& Contini(2004)]{pilyugin04} Pilyugin, 
  L.~S., V{\'{\i}}lchez, J.~M., \& Contini, T.\ 2004, \aap, 425, 849
\bibitem[Puche \& Carignan(1988)]{puc88} Puche, D., \& Carignan,
  C. 1988, \aj, 95, 1025
\bibitem[Rizzi et al.(2007)]{riz07} Rizzi, L., Tully, R.~B., 
Makarov, D., et al.\ 2007, \apj, 661, 815
\bibitem[Rousseau et al.(1978)]{rousseau78} Rousseau, J., Martin, N., Pr{\'e}vot, L., et al.\ 1978, \aaps, 31, 243
\bibitem[Russell \& Dopita(1990)]{russell90} Russell, S.~C., \& Dopita, M.~A.\ 1990, \apjs, 74, 93
\bibitem[Ryder et al.(1993)]{ryd93} Ryder, S., 
Staveley-Smith, L., Dopita, M., et al.\ 1993, \apj, 416, 167
\bibitem[Schlegel, Finkbeiner, \& Davis(1998)]{sch98} Schlegel, D.~J.,
  Finkbeiner, D.~P., \& Davis, M. 1998, \apj, 500, 525
\bibitem[Smartt et al.(2004)]{sma04} Smartt, S.~J., Maund, 
J.~R., Hendry, M.~A., et al.\ 2004, Science, 303, 499
\bibitem[Smartt et al.(2009)]{sma09} Smartt, S.~J., Eldridge, J.~J.,
  Crockett, R.~M., \& Maund, J.~R.  2009, \mnras, 395, 1409
\bibitem[Smith et al.(2011)]{smith11} Smith, N., Li, W., Silverman, J. M.,
  Ganeshalingam, M., \& Filippenko, A. V. 2011, \mnras, {\bf 415}, 773
\bibitem[Sonneborn, Altner, \& Kirshner(1987)]{son87} Sonneborn, G.,
  Altner, B., \& Kirshner, R.~P. 1987, \apjl, 323, L35
\bibitem[Stetson(1987)]{ste87} Stetson, P.~B. 1987, \pasp, 99, 191
\bibitem[Storchi-Bergmann, Calzetti, \& Kinney(1994)]{sto94}
  Storchi-Bergmann, T., Calzetti, D., \& Kinney, A.~L. 1994, \apj,
  429, 572
\bibitem[Terlevich et al.(1991)]{ter91} Terlevich, R., Melnick, J.,
  Masegosa, J., Moles, M., \& Copetti, M.~V.~F. 1991, \aap, 91, 285
\bibitem[Tonry et al.(2001)]{tonry01} Tonry, J.~L., Dressler, 
A., Blakeslee, J.~P., et al.\ 2001, \apj, 546, 681
\bibitem[Tully(1988)]{tully88} Tully, R.~B.\ 1988, Nearby Galaxies
  Catalog, Cambridge and New York, Cambridge University Press, 221 p.
\bibitem[Turatto et al.(1998)]{tur98} Turatto, M., Mazzali, 
P.~A., Young, T.~R., et al.\ 1998, \apjl, 498, L129
\bibitem[Van Dyk(2005)]{van05} Van Dyk, S.~D. 2005, The Fate of the
  Most Massive Stars, 332, 47
\bibitem[Van Dyk \& Matheson(2011)]{van11} Van Dyk, S.~D., \& Matheson, T. 2011, ApJ, 
in press
\bibitem[Van Dyk, Li, \& Filippenko(2003)]{van03} Van Dyk, S.~D., Li,
  W., \& Filippenko, A.~V. 2003, \pasp, 115, 1289
\bibitem[Van Dyk et al.(2002)]{van02} Van Dyk, S.~D., 
Garnavich, P.~M., Filippenko, A.~V., et al.\ 2002, \pasp, 114, 1322
\bibitem[Welch et al.(2007)]{wel07} Welch, D.~L., Clayton, 
G.~C., Campbell, A., et al.\ 2007, \apj, 669, 525
\bibitem[Woosley \& Weaver(1986)]{woo86} Woosley, S. E., \& Weaver,
  T. A. 1986, \araa, 24, 205
\bibitem[Zampieri et al.(2003)]{zam03} Zampieri, L., 
Pastorello, A., Turatto, M., et al.\ 2003, \mnras, 338, 711
\bibitem[Zwicky(1964)]{zwi64} Zwicky, F. 1964, \apj, 139, 514

\end{thebibliography}
\end{document}